\documentclass{nature}

\usepackage{amsmath,amssymb}
\usepackage{epsfig}
\usepackage{subfigure}
\usepackage{graphicx}
\usepackage{todonotes}
\usepackage{dcolumn}
\usepackage{bm}
\usepackage[colorlinks,linkcolor=blue,urlcolor=magenta,citecolor=blue,hyperindex,CJKbookmarks]{hyperref}
\usepackage{float}
\usepackage{hyperref}
\usepackage{comment}
\usepackage{color}
\usepackage{lineno}
\usepackage{soul}

\newcommand{\onlinecite}[1]{\hspace{-1 ex} \nocite{#1}\citenum{#1}}

\newcommand{\sio}{{Sr$_2$IrO$_4$}}

\title{Evolution of electronic and magnetic properties of Sr$_2$IrO$_4$ under strain}

\author{Ekaterina M. P{\"a}rschke$^{1,	*}$, Wei-Chih Chen$^{2}$,
Rajyavardhan Ray$^{3,4,5}$ \& Cheng-Chien Chen$^{2}$}

\makeatletter
\let\saved@includegraphics\includegraphics
\AtBeginDocument{\let\includegraphics\saved@includegraphics}
\renewenvironment*{figure}{\@float{figure}}{\end@float}
\makeatother

\begin{document}
\maketitle

\begin{affiliations}

\item Institute of Science and Technology Austria, Am Campus 1, 3400 Klosterneuburg, Austria
\item Department of Physics, University of Alabama at Birmingham, Birmingham, Alabama 35294, USA
\item Leibniz IFW Dresden, Helmholtzstr. 20, 01069 Dresden, Germany
\item Dresden Center for Computational Material Science (DCMS), TU Dresden, 01062 Dresden, Germany
\item Department of Physics, Birla Institute of Technology Mesra,
      Ranchi, India 835215
\end{affiliations}

\noindent
Email: ekaterina.paerschke@ist.ac.at

\bigskip

\begin{abstract}

     Motivated by properties-controlling potential of the strain, we investigate strain dependence of structure, electronic and magnetic
     properties of {\sio} using complementary
     theoretical tools: {\it ab-initio} calculations,
     analytical approaches (rigid octahedra picture, 
     Slater-Koster integrals), and extended $t-{\mathcal{J}}$ model. 
     We find that strain affects both Ir-Ir distance and Ir-O-Ir angle, and the rigid octahedra picture is not relevant. Second, we find fundamentally different behavior for compressive and tensile
     strain. 
     One remarkable feature is the formation of two subsets of
     bond- and orbital- dependent carriers, a compass-like model, 
     under compression. This originates from the strain-induced renormalization of the Ir-O-Ir superexchange and O on-site energy.
     We also show that under compressive (tensile) strain, Fermi surface becomes highly dispersive (relatively flat).
     Already at a tensile strain of $1.5\%$, we observe spectral weight redistribution, with the low-energy band acquiring almost purely singlet character.
     These results can be directly compared with future experiments. 
     
\end{abstract}
%\pacs{}

\section*{Introduction} 

Exploring the physics of quasi-two-dimensional (2D) spin-orbit Mott
insulators can
help to understand high-temperature superconductivity as well as
the general interplay of spin-orbit coupling, Hund's, and Coulomb
interactions. In particular, a lot of studies have been devoted to the
quasi-2D iridates Sr$_2$IrO$_4$ and
Ba$_2$IrO$_4$.~\cite{Kim2008,Jackeli2009,CaoReview2018}
Iridates show eminent similarities to the cuprate family of
high-temperature superconductors, both in structure and low-energy
physics, and were expected to become superconducting upon doping.  
However, so far no superconductivity has been reported in iridates. 

In general, {\sio} behavior often deviates from theoretical predictions. 
For example, Mott insulators normally become metallic at high
enough pressure as the unit cell becomes smaller 
and the bands broaden. This is also 
true for spin-orbit coupled Mott insulators, such as ruthenates.~\cite{Snow2002,CaoReview2018} 
In Sr$_2$IrO$_4$, resistance indeed
decreases until the pressure of around 25-30 GPa (which according to Ref. ~\onlinecite{Haskel2012}, corresponds approximately to a strain of -4\%),~\cite{Haskel2012} or, according to a very recent study~\onlinecite{ChunhuaChen2019}, 32-38 GPa (-5.1\% strain). Then, however, resistance
starts to increase, showing a peculiar U-shaped dependency and
persisting insulating behavior up to at least 185
GPa.~\cite{ChunhuaChen2019} So far, no metallization in Sr$_2$IrO$_4$ or
other iridates (Sr$_3$Ir$_2$O$_7$~\cite{Ding2016,JZhang2018},
BaIrO$_3$,~\cite{Korneta2010,CaoReview2018} etc) has been observed at
pressures up to 40-185 GPa.~\cite{CaoReview2018,ChunhuaChen2019}
Moreover, there is also surprisingly little correlation between the
insulating behavior and magnetism~\cite{GCao1998} as the latter
disappears at around 20 GPa (roughly -2.9 \% strain) in Sr$_2$IrO$_4$~\cite{Haskel2012, HaskelPRL2021} and 14.4
GPa (roughly -2.1 \% strain) in Sr$_3$Ir$_2$O$_7$,~\cite{JZhang2018} without the onset of a
metal-insulator transition.

Furthermore, iridates emerge as a good functional playground for
manipulation of the magnetic and electronic properties, which is
an exciting goal both fundamentally and
practically.~\cite{RayanSerraoPRB2013,CaoReview2018} Iridium-based
heterostructures and superlattices have therefore emerged as a whole new
field very recently.~\cite{NicholsNatCom2016,DiYiPNAS2016,OkamotoNanoLet2017,GruenewaldAdvMat2017,JWKimPRB2018,WenNJP2019,MeyersSciRep2019,DasaJMatChemC2019,Mohanta2019}
Strain and pressure in particular %and doping 
are powerful tools on hand to control the magnetic properties of the material.
It has been shown that misfit strain can directly control dispersion of magnetic excitations in {\sio},
~\cite{LupascuPRL2014,Paris2020,SeoPRB2019} as well as transport
properties.~\cite{SouriJAP2019} A shift of the two-magnon
Raman peak to higher energies was observed under tensile
strain,~\cite{SeoPRB2019} albeit much weaker than the shift observed in
the canonical Mott-Hubbard
insulator K$_2$NiF$_4$ and cuprates like
Bi$_{1.98}$Sr$_{2.06}$Y$_{0.68}$Cu$_2$O$_{8+\delta}$.~\cite{CukPRL2008}

In Ref. [\onlinecite{Paris2020}], the authors used resonant inelastic scattering (RIXS)
to show that magnetic dispersion in {\sio} is strongly affected by strain.
In particular, the contribution of the second and third nearest-neighbor
(NN) exchange was suppressed (enhanced) upon tensile (compressive)
strain. The tensile strain was shown to drive the system closer to a
shorter-range first-NN only Heisenberg limit, with only little magnon
branch softening left at ($\rm{\pi}/2,\rm{\pi}/2$) already upon the tensile strain of
$+2\%$. Upon compressive strain, the energy of ($\mathrm{\pi},0$) magnon was
shown to increase.~\cite{Paris2020,LupascuPRL2014}

A clear understanding of the electronic and magnetic properties of
iridates and their evolution with strain is, therefore, of interest not only from a fundamental point of view but also
for applications.~\cite{RayanSerraoPRB2013,CaoReview2018} Unveiling the
details of the interplay of lattice, magnetic, and other degrees of
freedom in {\sio} is needed to understand the recently observed electrical
control of octahedra rotation~\cite{CaoPRL2018,YePRB2020} and the much-debated strong magnetoelastic coupling.~\cite{SonNPJQM2019,DashwoodPRB2019, Liu2019, SamanthaAPL2019}
Currently, a clear understanding of neither how exactly nor by which mechanism do
superexchange and hopping parameters in {\sio} change with strain is
available, not even on a
phenomenological level. One of the interesting questions is whether
the change in electronic and magnetic properties upon the strain is mostly
associated with bond length change, as argued in e.g. Ref.~[\onlinecite{LupascuPRL2014}], or the change of the in-plane
rotation angles $\theta$ of the oxygen octahedra (see
Fig.~\ref{cartoon}).\cite{NicholsAPL2013} 

Studying the behavior of iridates under strain and pressure is a
demanding task not only experimentally, but also theoretically. On one
hand, iridates are strongly correlated Mott insulators,~\cite{Kim2008}
so one needs to resort to theoretical methods where correlations are
treated non-perturbatively, employing effective descriptions like Hubbard
or Heisenberg models. On the other hand, microscopic changes of orbitals, their
overlap, and structural changes are essential to understand the behavior of a
crystal under strain,~\cite{BhandariNJP2019,ZaitsevJPCM2020,ChoiPRB2020,EngstromPRB2021,NicholsonComNat2021} so {\it ab-initio} methods are
demanded. 
Another difficulty is that as one eventually approaches a possible
metal-insulator transition at high pressure and/or strain, effective
models, such as the Heisenberg superexchange model, fail.

In this paper, we focus on the effect of strain and combine various
complementary theoretical tools to provide a comprehensive analysis of how the magnetic properties are affected by strain.
For different (compressive and tensile) strain
values, we use density functional theory (DFT) based {\it ab-initio} calculations to access
microscopic changes in the crystal structure, and study the
corresponding changes in the electronic properties through
Wannierization of the scalar-relativistic DFT bandstructure obtained within the generalized
gradient approximation (GGA).\cite{PerdewPRL1996} Subsequently, we
solve an extended $t-{\mathcal{J}}$ model within the  
self-consistent Bohr approximation (SCBA) to obtain the angle-resolved photoemission spectra (ARPES) and study the strain-controlled evolution of the Fermi surface. Realistic
values of the input parameters for these
calculations were used: the hopping
parameters were obtained from the DFT calculations, while the
extended-range exchange couplings were obtained by direct comparison
to the magnon
dispersion measured with RIXS.
In this way, the presented analysis contains no free parameters
apart from an overall constant energy shift (chemical potential) in the
SCBA calculations.
\section*{Results and Discussion}

\subsection{Evolution of hopping parameters under strain.}

\begin{figure}
	\centering
	\includegraphics[width=1.0\linewidth]{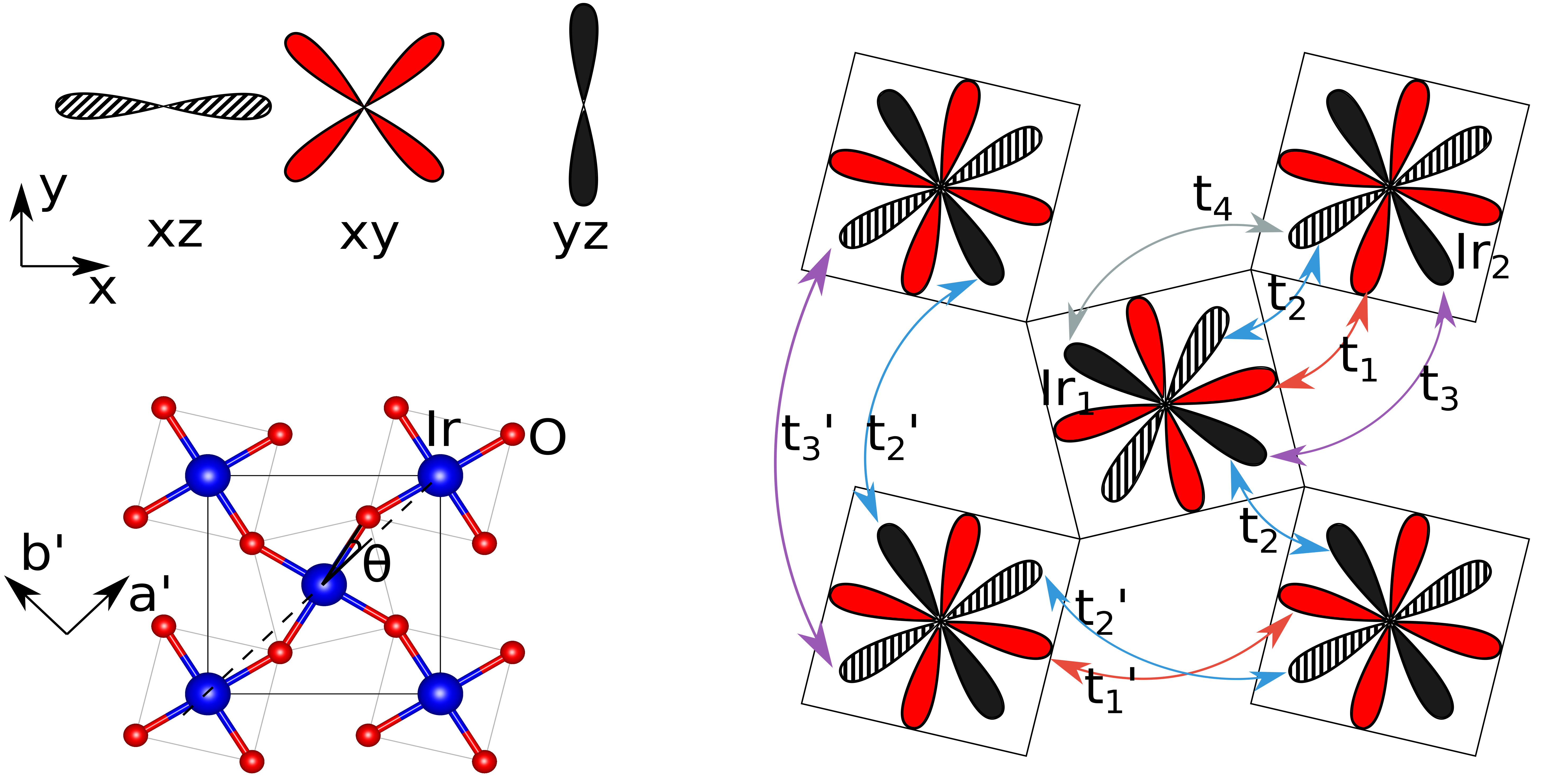}
	\caption{\textbf{In-plane structure of {\sio} and the notation for orbital overlap parameters.} For nearest neighbors (NN), interorbital overlap parameters are denoted $t_1$, $t_2$, $t_3$ and
		intraorbital $t_4$, and for second nearest neighbors (2NN) we consider interorbital overlap only: $t_1'$,
		$t_2'$, $t_3'$. Orbital overlap between third neighbors is defined in the same way, not shown. Due to the symmetry considerations, the overlap of out-of-plane $t_{\mathrm{2g}}$ orbitals is anisotropic: $t_2$ describes hopping between $xz$($yz$) along $a'$ ($b'$) axis. The O-Ir-Ir angle $\theta$ characterizes the
		in-plane octahedra rotation as shown.
		\label{cartoon}}
\end{figure}

\begin{figure}
	\centering
	\includegraphics[width=0.48\linewidth]{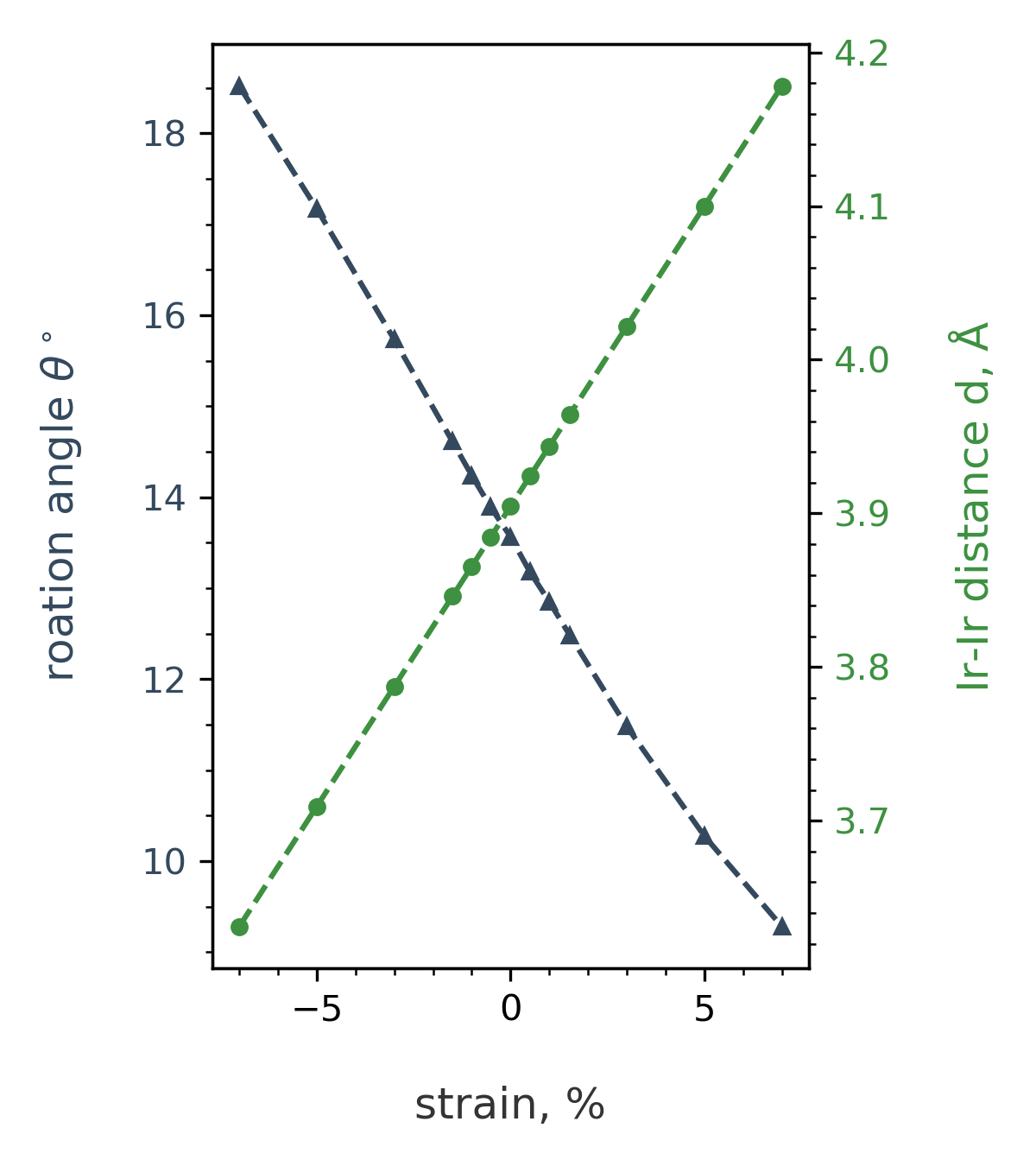}\label{thetadDFT}
	\llap{\parbox[b]{3.00in}{\textbf{a}\\\rule{0ex}{3.2in}}}
	\includegraphics[width=0.48\linewidth]{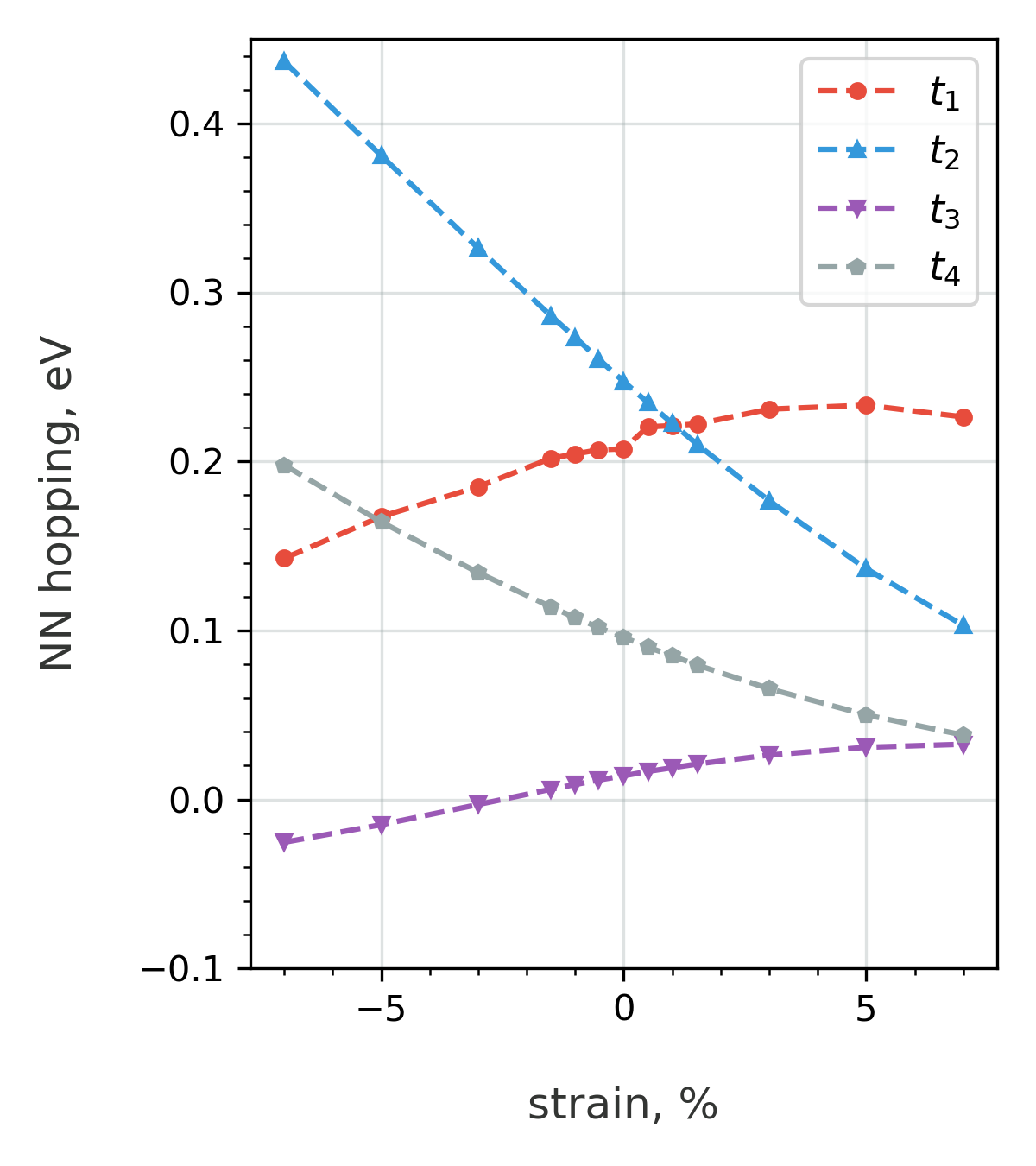}\label{tNNDFT}
	\llap{\parbox[b]{3.0in}{\textbf{b}\\\rule{0ex}{3.2in}}}
	\caption{\textbf{Dependency of the octahedra rotation angle $\theta$, Ir-Ir distance $d$ and NN hopping parameters on strain.} \textbf{a} The in-plane IrO$_6$ octahedral rotation angle $\theta$
		and Ir-Ir distance $d$ as functions of strain, obtained with
		full structure relaxation in DFT calculations (see Methods). \textbf{b}
		NN hopping parameters for different values of strain, see
		Fig.\ref{cartoon} for notation.
	}
	\label{t_and_theta_d_under_strain}
\end{figure}

{\sio} shows an in-plane staggered octahedra
rotation characterized by a single parameter: $\measuredangle {\rm O}-{\rm
	Ir}-{\rm Ir}$ angle denoted by $\theta$ in Fig.~\ref{cartoon}.
Under ambient conditions, the octahedral rotation is found to be
$\theta \approx 13.6^{\circ}$ for the relaxed
structure, which is close to the reported experimentally value of $11.8^{\circ}$.~\cite{Boseggia2013} 
The epitaxial strain on iridates then affects not only the distance between the Ir atoms but also the Ir-O-Ir
bond angle, as can be seen in Fig.~\ref{t_and_theta_d_under_strain}a.
The in-plane octahedra rotation angle $\theta$, obtained using DFT (see
Methods for details), monotonically increases (decreases) upon
compressive (tensile) strain in the studied range of -7.5$\%$ to
7.5$\%$, where negative strains correspond to compression.

%%%%%%%%%%%%%%%% FIG 1 %%%%%%%%%%%%%
%%%%%%%%%%%%%%%% FIG 1 %%%%%%%%%%%%%

To ascertain the influence of structural changes on the electronic properties, we study the evolution of Wannier tight-binding model hoppings derived from DFT (see Methods for details), as a function of strain (Fig.~\ref{t_and_theta_d_under_strain}b).
The notations for the hoppings are shown in Fig 1: the intraorbital hoppings between $xy$ orbitals along $a'$ or $b'$ axes is denoted as $t_1$, between $xz$($yz$) along $a'$ ($b'$) axis as $t_2$, and between $xz$($yz$) along $b'$
($a'$) as $t_3$. The interorbital hopping between $yz$ and $xz$ orbitals is denoted as $t_4$, all other interorbital hoppings are negligible. Further neighbor interorbital hoppings are denoted as $t_i'$ and are shown in Fig. 1.  

Upon compression, direction-dependent hopping parameter $t_2$ is increasing, but surprisingly, $t_1$ is decreasing (Fig.~\ref{t_and_theta_d_under_strain}b).
This emerging anisotropy in hopping parameters is interesting, as
$t_2$ hopping describes the propagation of an electron with $xz$ ($yz$)
orbital character along only one axis, $a'$ ($b'$), whereas $t_1$ allows
an $xy$ electron to hop in both directions. We thus see that upon
compressive strain, the system favors the separation of the entire Fermi
sea into two Fermi seas with bond-dependent propagation  
($xz$ carriers which can only propagate along $a'$, and $yz$ carriers which can only propagate along $b'$) 
and suppression of the bond-independent and thus truly two-dimensional $xy$ carriers. 
This compass-model-like\cite{Nussinov2015} propagation is quite unusual and could cause the formation of charge density wave.

Upon tensile strain, $t_1$ is nearly
independent of the strain value and is the dominant hopping, while $t_2$ decreases steadily (Fig.~\ref{t_and_theta_d_under_strain}b). Different behavior of $t_1$ upon compressive and tensile strain reflects
the change of Fermi surface topology between compressive and tensile
strain.

It is also interesting to note that the smallest hopping parameter
$t_3$, describing the hopping between almost parallel $d$-orbitals
with very small overlap goes to zero around -3$\%$, which
corresponds to compression of $\sim 20\,$GPa, not too far from the
value of resistivity minimum under pressure.~\cite{Haskel2012,ChunhuaChen2019} 

To disentangle the contribution of inter-atomic distance $d$ and
the octahedral rotation $\theta$ to the hopping parameter trends with
strain, we employ the analytical approaches of Glazer and
Slater-Koster.
The Glazer picture \cite{Glazer1972} is often used in rigid
    octahedra approximation whereby the 
main effect of the modest strain is assumed to be the
change of the in-plane rotation angle $\theta$. 
However, as detailed in Supplementary Figure 2, the trends obtained within the Glazer picture disagree with the DFT results in
Fig.\ref{t_and_theta_d_under_strain}b, and even contradict them
in rigid octahedra approximation. Therefore,
the Glazer picture has limited applicability for iridates, and rigid octahedra approximation is improper.

%%%%%%%%%%%%%%%% FIG 2 %%%%%%%%%%%%%
%%%%%%%%%%%%%%%% FIG 2 %%%%%%%%%%%%%

We then proceed with a more specific orbital-resolved
Slater-Koster-integrals-based approach.~\cite{HarrisonBook2004,Slater1954} %To this end, 
Slater-Koster integrals are hybridization matrix elements $E$ between atomic $d$-states on neighboring atoms obtained via integrating over relevant spherical harmonics. The resulting interatomic matrix elements $E$ are proportional to the $d$-wave functions overlap and can be expressed via cubic
harmonic matrix elements $V_{\mathrm{d d \sigma}}$,$V_{\mathrm{d} \mathrm{d} \mathrm{\pi}}$, $V_{\mathrm{d d\delta}}$ for a known bond direction ${l,m,n}$ as tabulated in Slater-Koster tables.\cite{HarrisonBook2004,Slater1954}
In {\sio}, we also need to account
for the rotation of the $d$ orbitals within the $t_{\mathrm{2g}}$ sector due to
the in-plane octahedral rotation.\cite{WangSenthil2011} Therefore, we decompose
the rotated $d$ orbital in the basis of
non-rotated $d$ orbitals before evaluating the Slater-Koster matrix
elements. 
For example,
the hybridization matrix element $\tilde{E}$ between the two rotated  NN $xy$ orbitals can be obtained as a superposition of hybridization matrix elements $E$ of non-rotated $xy$ and $x^2-y^2$ orbitals obtained as:
\begin{eqnarray}
	 \tilde{E}_{\mathrm{xy,xy}}=\cos^2(2\theta)E_{\mathrm{xy,xy}}-\sin^2(2\theta)E_{\mathrm{x^2-y^2,x^2-y^2}},	
	 \label{Exyxy}
\end{eqnarray}
where $\theta$ is the in-plane rotation of the IrO$_6$ octahedra (see Fig.\ref{cartoon}). Similarly, for the overlap between
the rotated $xz$($yz$) orbitals along the $a'$ direction, we get
\begin{eqnarray}
	\tilde{E}_{\mathrm{xz,xz (yz,yz)}}=\cos^2(\theta)E_{\mathrm{xz,xz (yz,yz)}}-\sin^2(\theta)E_{\mathrm{yz,yz (xz,xz)}},
	\label{Exzxz}
\end{eqnarray}
and for interatomic interorbital overlap along $x$:
\begin{eqnarray}
 	\tilde{E}_{\mathrm{yz,xz}}=E_{\mathrm{yz,xz}}+\sin\theta\cos\theta(E_{\mathrm{xz,xz}}+E_{\mathrm{yz,yz}}).
 	\label{Exzyz}
\end{eqnarray}

\begin{figure}
	\centering
	\includegraphics[width=0.3525\linewidth]{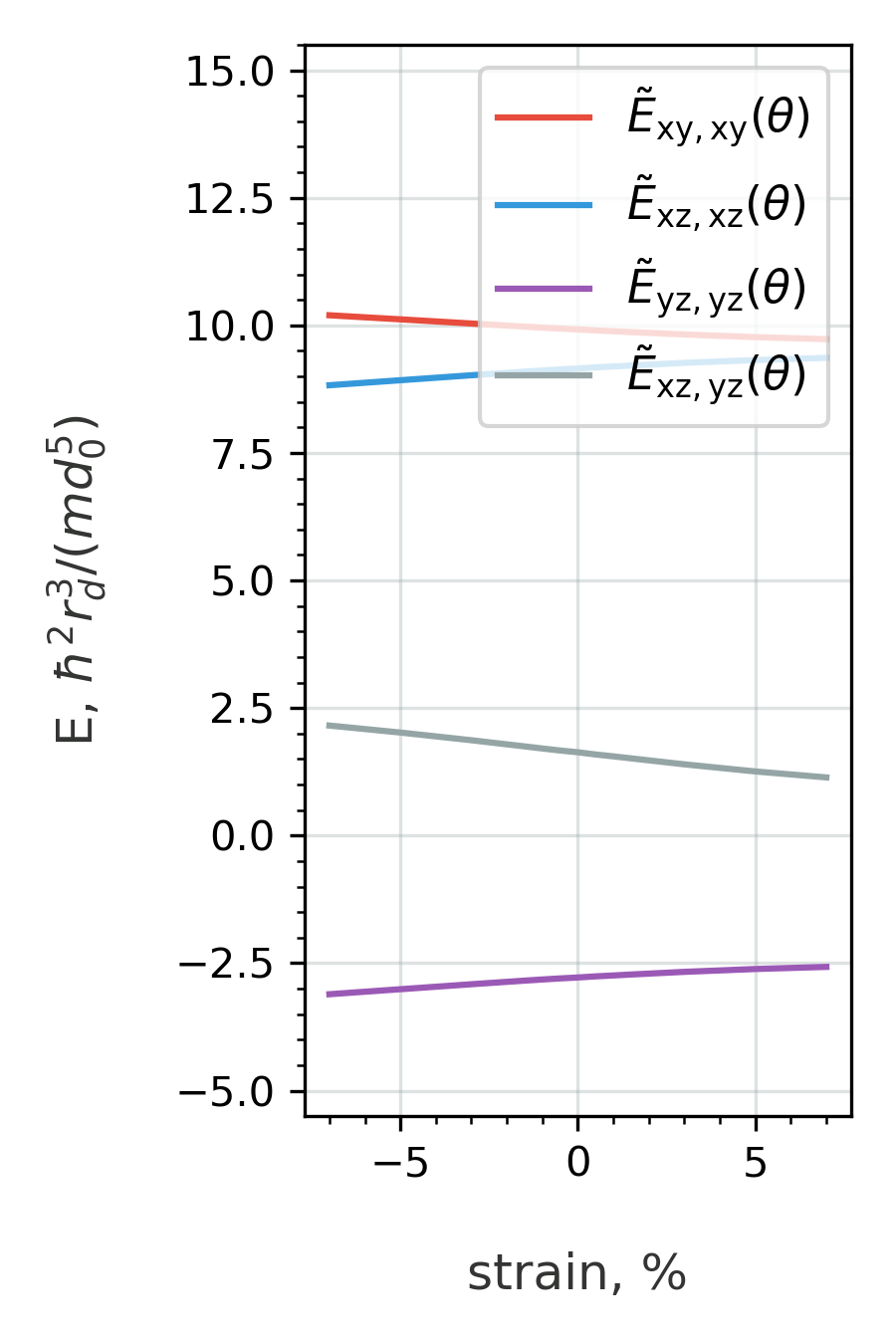}\label{ENNSLaterKostertheta}
	\llap{\parbox[b]{2.2in}{\textbf{a}\\\rule{0ex}{3.2in}}}
	\includegraphics[width=0.29375\linewidth]{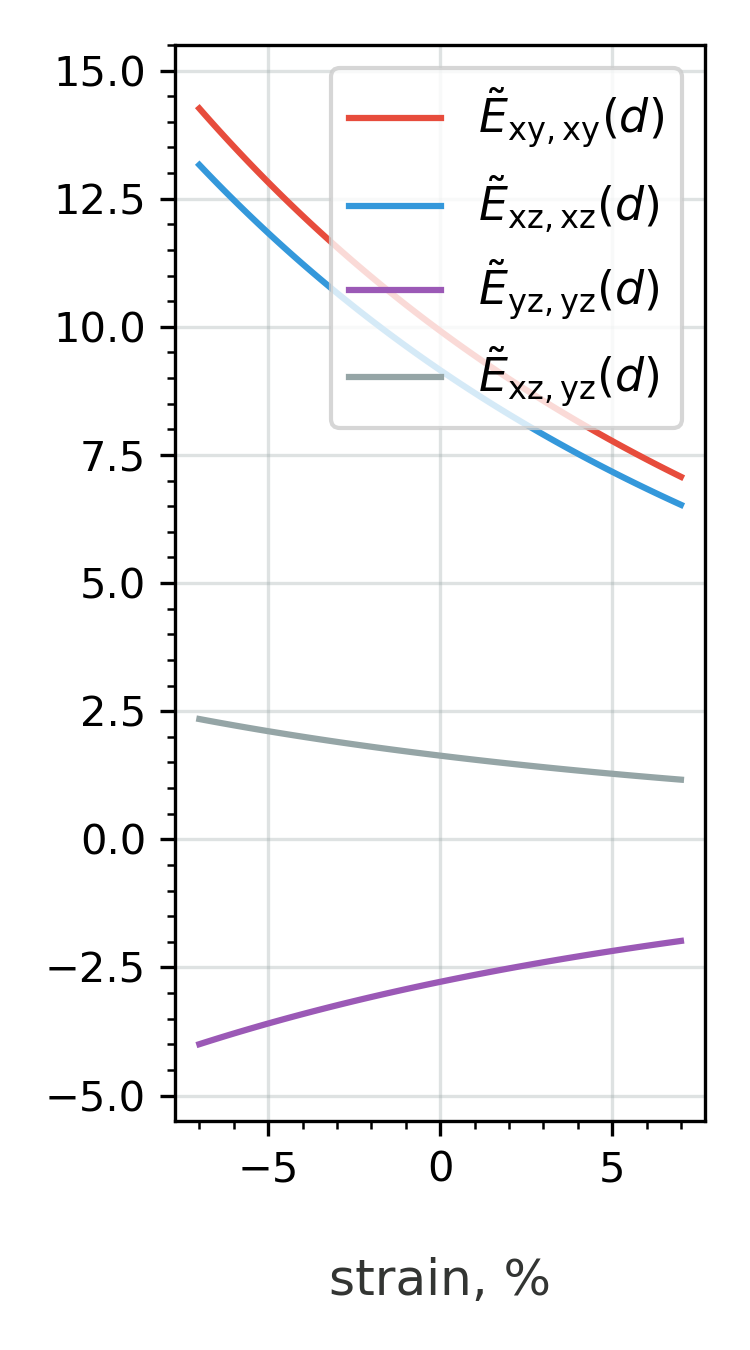}\label{ENNSLaterKosterd}
	\llap{\parbox[b]{2.0in}{\textbf{b}\\\rule{0ex}{3.2in}}}
	\includegraphics[width=0.29375\linewidth]{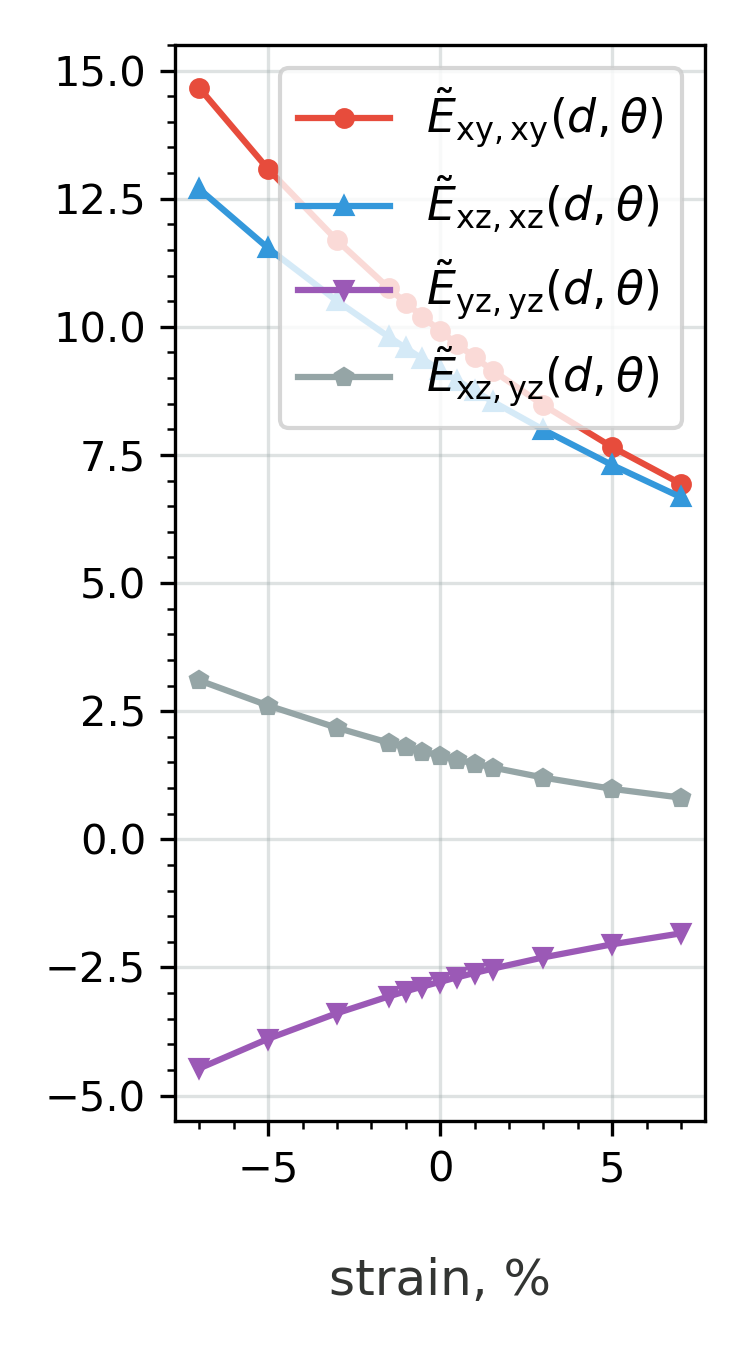}\label{ENNSLaterKoster}
	\llap{\parbox[b]{2.0in}{\textbf{c}\\\rule{0ex}{3.2in}}}
	\caption{\textbf{Dependency of Slater-Koster Ir-Ir integrals on strain.} Slater-Koster integrals for rotated NN orbitals $\tilde{E}$ as a function of \textbf{a} octahedra in-plane rotation angle $\theta$, \textbf{b} Ir-Ir distance $d$, \textbf{c} both $\theta$ and $d$ (calculated using values of $\theta$ and $d$ extracted from DFT 
		shown in Fig.~\ref{t_and_theta_d_under_strain}a). 
		\label{E_fct_of_d_theta_full}}
\end{figure}

Figure ~\ref{E_fct_of_d_theta_full} shows the resulting hybridization matrix elements
$\tilde{E}$ as a function of the in-plane octahedral rotation $\theta$,
the Ir-Ir distance $d$, as well as both the parameters (see Fig.
\ref{t_and_theta_d_under_strain}a). We find that at least in the Slater-Koster approximation,
accounting for the change of the distance $d$ alone
(Fig.~\ref{E_fct_of_d_theta_full}b) can provide a better approximation to a
full dependency of matrix elements $\tilde{E}$ on strain
(Fig.~\ref{E_fct_of_d_theta_full}c) then accounting for the change of bond angle $\theta$. 
This is also consistent with the quantum chemistry
study.~\cite{LupascuPRL2014}

However, not all trends obtained from the DFT calculations are well reproduced:   
the hopping parameter $E_{\mathcal{xy,xy}}$ is 
increasing under compressive strain (Fig.
\ref{E_fct_of_d_theta_full}c), unlike the $t_1$ hopping extracted from DFT
(Fig.~\ref{t_and_theta_d_under_strain}b).
To address this, we also consider the O-mediated indirect
Ir$-$O$-$Ir hoppings.

%%%%%%%%%%%%%%%% FIG 3%%%%%%%%%%%%%
%%%%%%%%%%%%%%%% FIG 3 %%%%%%%%%%%%%

The indirect oxygen-mediated overlap between the two rotated NN $xy$ orbitals can be calculated as a sum of the hopping integrals between two Ir atoms via $\alpha=p_{\mathrm{x}}$, $p_{\mathrm{y}}$ orbitals of the oxygen, $\tilde{E}_{\mathrm{xy,O,xy}}=\sum_{\alpha =p_{\mathrm{x}},p_{\mathrm{y}}}{E_{\mathrm{xy},\alpha,\mathrm{xy}}}$. The hopping integral is calculated as
\begin{eqnarray}
	\tilde{E}_{\mathrm{xy},\alpha,\mathrm{xy}}&=(\cos(2\theta)E_{\alpha,\mathrm{xy}}^{-l,m,n}+\sin(2\theta)E_{\alpha,\mathrm{x^2-y^2}}^{-l,m,n}) \nonumber \\
	&\times(\cos(2\theta)E_{\alpha,\mathrm{xy}}^{l,m,n}-\sin(2\theta)E_{\alpha,\mathrm{x^2-y^2}}^{l,m,n})/\Delta_{\mathrm{pd}},
	\label{ExyOxy}
\end{eqnarray}
where $l=\cos{\theta}$, $m=\sin{\theta}$, $n=0$ are the directional
cosines of the vector from the oxygen O to the Ir
atom~\cite{HarrisonBook2004} along the $a'$-axis in the units of Ir-O
distance $d_{\rm Ir-O}= 0.5 d_0^{\rm Ir-Ir}/\cos{\theta}$ ($d_0^{\rm
Ir-Ir}$ is the distance between the NN Ir ions under ambient
conditions), and $E_{\alpha,\mathrm{xy}}$,
$E_{\alpha,\mathrm{xy}}$ are the $p$-$d$ Slater-Koster
integrals.~\cite{HarrisonBook2004}
We note that $l,m,n$ indexes were omitted for Eqs.(\ref{Exyxy}-\ref{Exzyz}), because for the $d-d$ overlap, Slater-Koster
integrals are quadratic in directional cosines,~\cite{HarrisonBook2004}
and simply $\{l,m,n\}=\{1,0,0\}$ for a pair of Ir atoms along the
$a'$ bond. For the indirect oxygen-mediated overlap, however, one has to account for the sign of the directional cosines. 
Moreover, for the indirect hopping, the Ir-O-Ir hopping has to be renormalized by the charge transfer energy $\Delta_{\mathrm{pd}}=E^{\mathrm{on-site}}_{\mathrm{xy}}-E^{\mathrm{on-site}}_{\alpha}$, the energy difference between corresponding Ir-$d$ and O-$p$ orbitals (see Supplementary Figure 3a). Surprisingly, the charge transfer energy $\Delta_{\mathrm{pd}}$ has a strongly non-linear dependency on the strain (see Supplementary Figure 3a).  

\begin{figure}
	\centering
	\includegraphics[width=0.3525\linewidth]{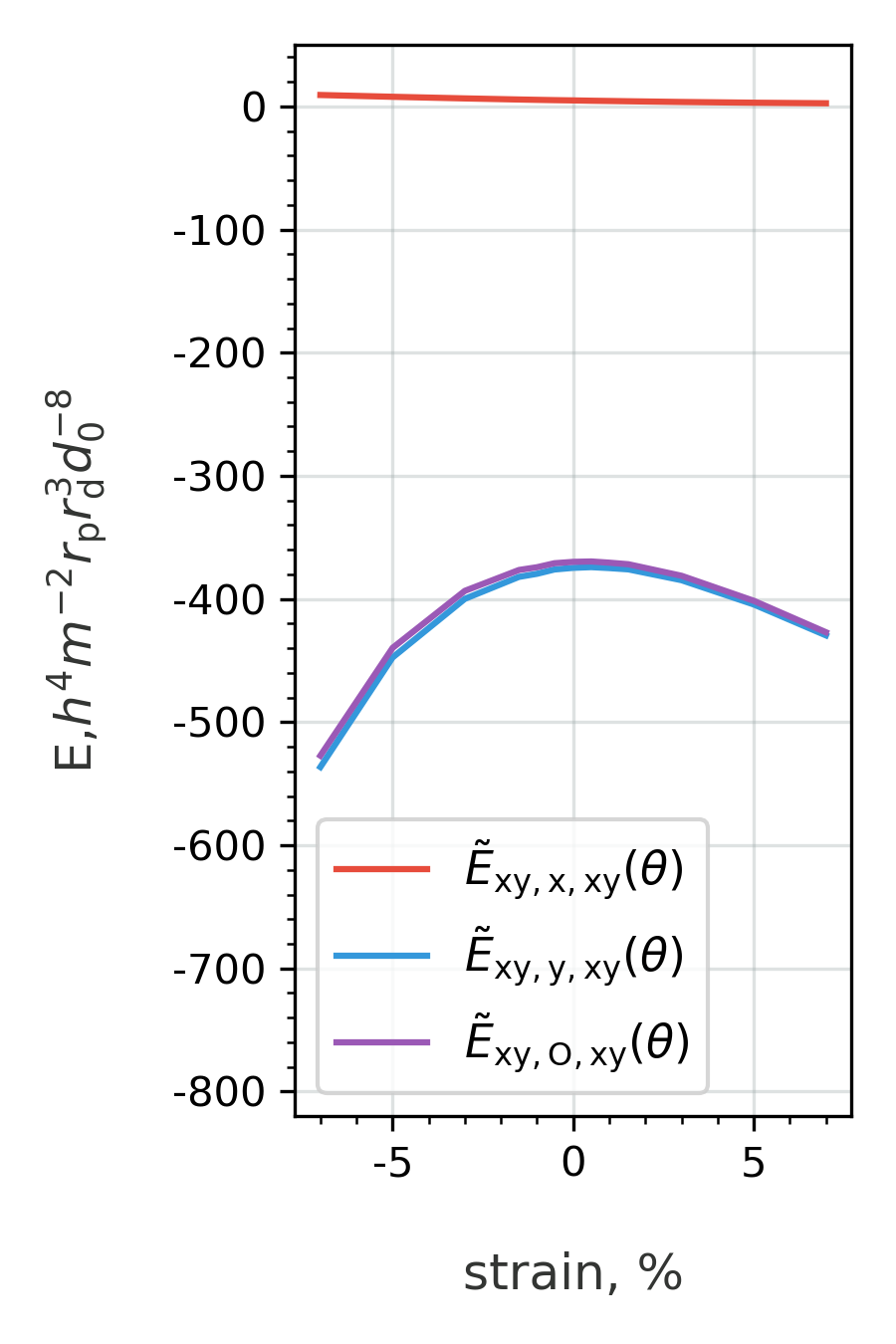}\label{EviaOSLaterKostertheta}
	\llap{\parbox[b]{2.2in}{\textbf{a}\\\rule{0ex}{3.2in}}}
	\includegraphics[width=0.29375\linewidth]{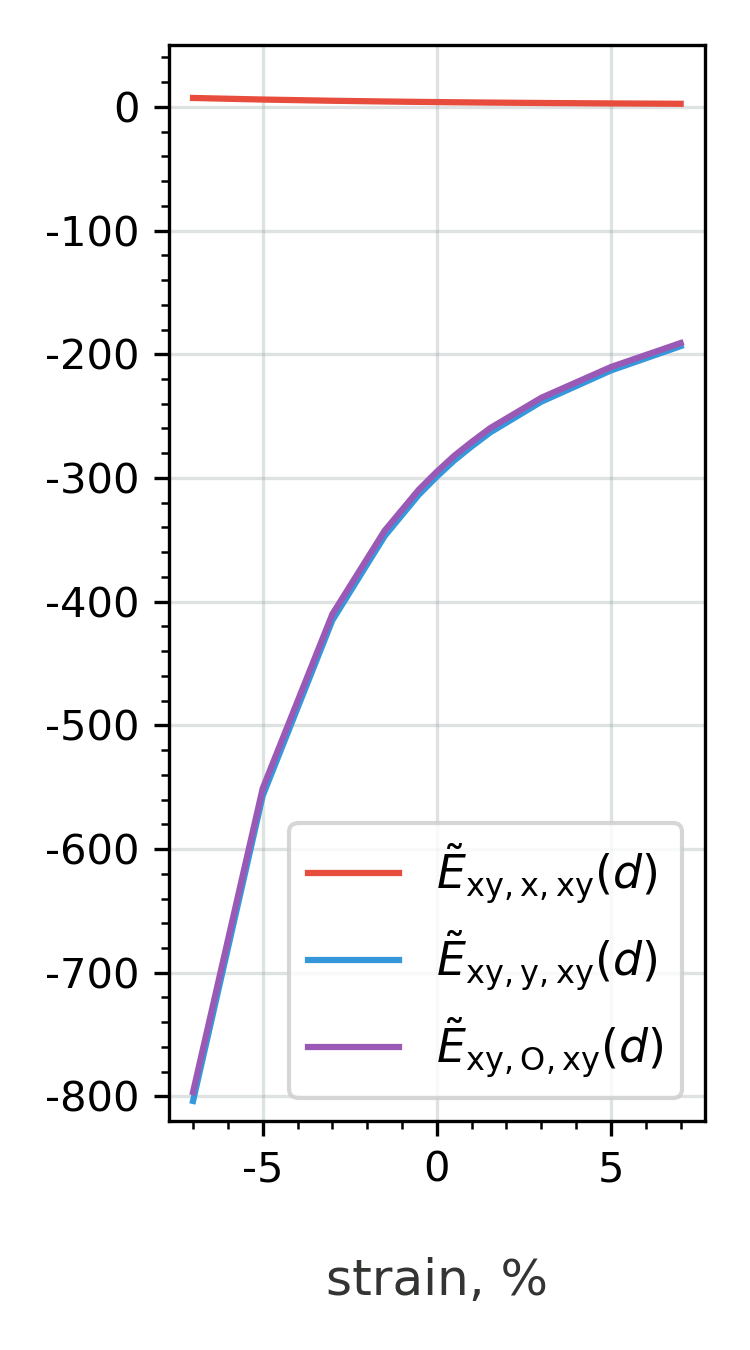}\label{EviaOSLaterKosterd}
	\llap{\parbox[b]{2.0in}{\textbf{b}\\\rule{0ex}{3.2in}}}
	\includegraphics[width=0.29375\linewidth]{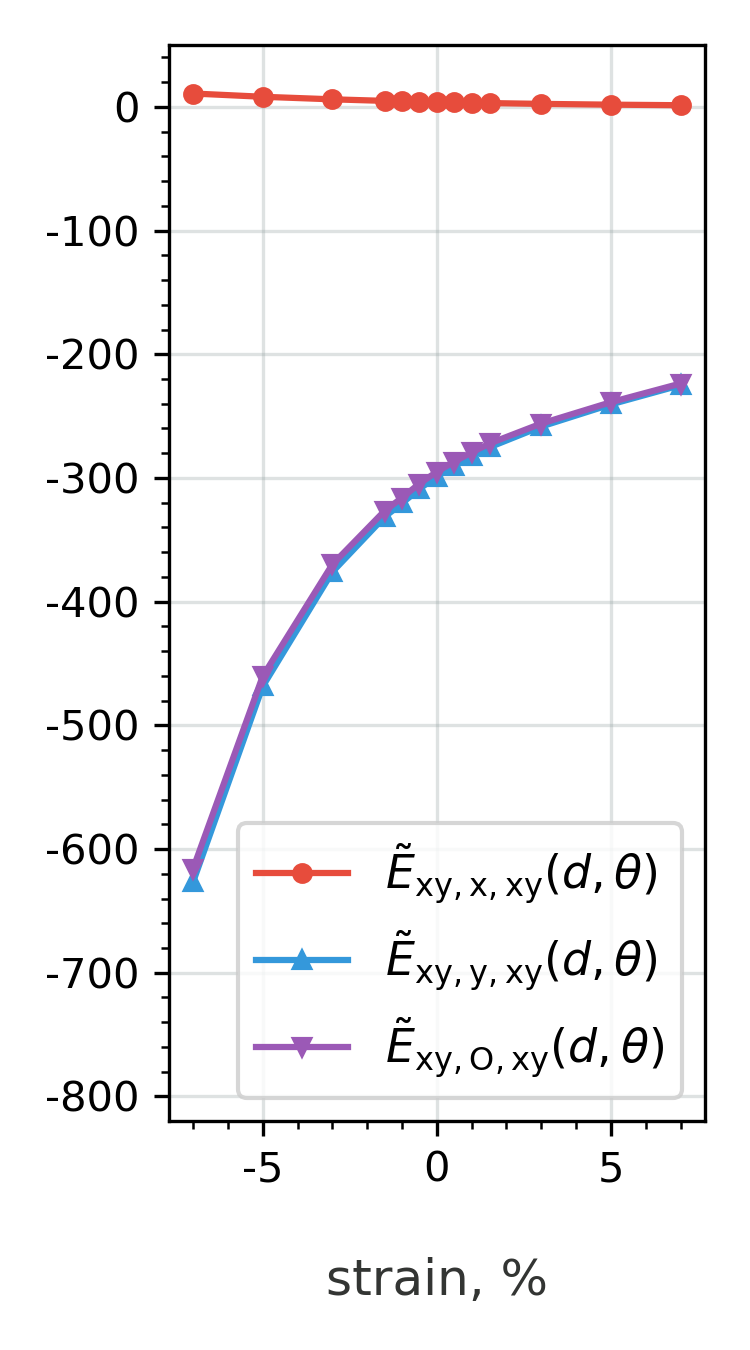}\label{EviaOSLaterKoster}
	\llap{\parbox[b]{2.0in}{\textbf{c}\\\rule{0ex}{3.2in}}}
	\caption{\textbf{Dependency of Slater-Koster Ir-O-Ir integrals on the strain.} Slater-Koster integrals for indirect hopping between rotated Ir NN orbitals $\tilde{E}$ via oxygen $p$ orbitals as a function of \textbf{a} octahedra in-plane rotation angle $\theta$, \textbf{b} Ir-Ir distance $d$, \textbf{c} both $\theta$ and $d$ (calculated using values of $\theta$ and $d$ extracted from DFT shown in Fig.~\ref{t_and_theta_d_under_strain}a). 
		\label{EviaO_full}}
\end{figure}

We plot the resulting indirect superexchange matrix elements
$\tilde{E}$ as a function of the in-plane octahedral rotation $\theta$,
Ir-Ir distance $d$ in Fig. \ref{EviaO_full}. Indeed, the indirect
Ir-Ir overlap decreases drastically under compressive strain, unlike the
direct $d-d$ overlap. Interestingly, this behavior is directly linked to both the change in the distance between the atoms as well as the angle $\theta$ describing octahedra rotation (see
Figs.\ref{EviaO_full}a,b). We also note that taking into account the
strain dependence of the iridium and oxygen on-site energies is crucial
to obtain correct trends. In fact, non-linear behavior of charge transfer energy $\Delta_{\mathrm{pd}}$ seems to be directly responsible for the non-linear strain dependency of the indirect oxygen-mediated hopping $\tilde{E}$. This suggests that the role of oxygens in the
low-energy physics of strained iridates and other transitional-metal
oxides might be underestimated and requires further investigation.

%%%%%%%%%%%%%%%% FIG 4 %%%%%%%%%%%%%
%%%%%%%%%%%%%%%% FIG 4 %%%%%%%%%%%%%

Accordingly, the contribution of the indirect orbital overlap should be small for $xz$ and $yz$ orbitals. 
Indeed, the indirect $yz$-$yz$ orbitals overlap along $a'$-axis is zero. 
The $xz$ orbitals hybridize with O-$p_\mathrm{z}$ orbital, however, this hybridization decreases with strain much slower than 
 for $xy$ orbitals, explaining the different behavior of $xy$ and $xz$ orbitals under compressive strain.

The fact that relative Ir-O hybridization is
directly responsible for the resulting suppression of $t_\mathrm{1}$ ($xy$-$xy$ hopping) under
compressive strain suggests an
electronic state crossover as the role of $xy$ orbitals in the
composite $J_\mathrm{eff}=1/2$ is decreasing. %Such 
Notably, a pressure-induced
phase transition was also suggested in a recent X-ray powder
diffraction study~\cite{Samanta2020} at pressures around
20 GPa, which should correspond to approximately $-3\%$ strain and
is in good agreement with our findings.

\subsection{Overlap of the spin-orbit coupled $J_\mathrm{eff}$ states.}
\begin{figure}
	\centering
	\includegraphics[width=0.5\linewidth]{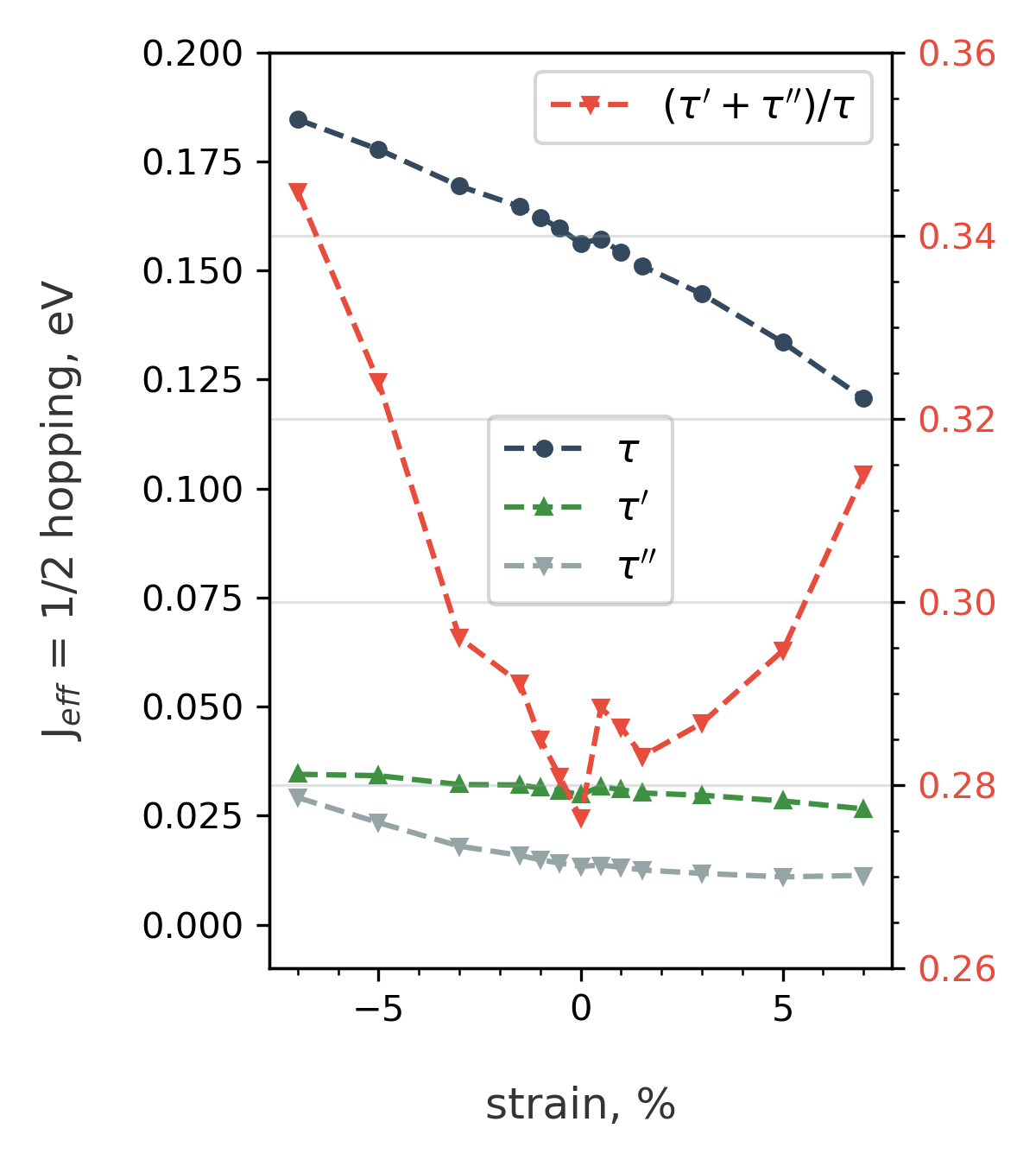}\label{tau}
	\llap{\parbox[b]{3.1in}{\textbf{a}\\\rule{0ex}{3.40in}}}
	\includegraphics[width=0.37\linewidth]{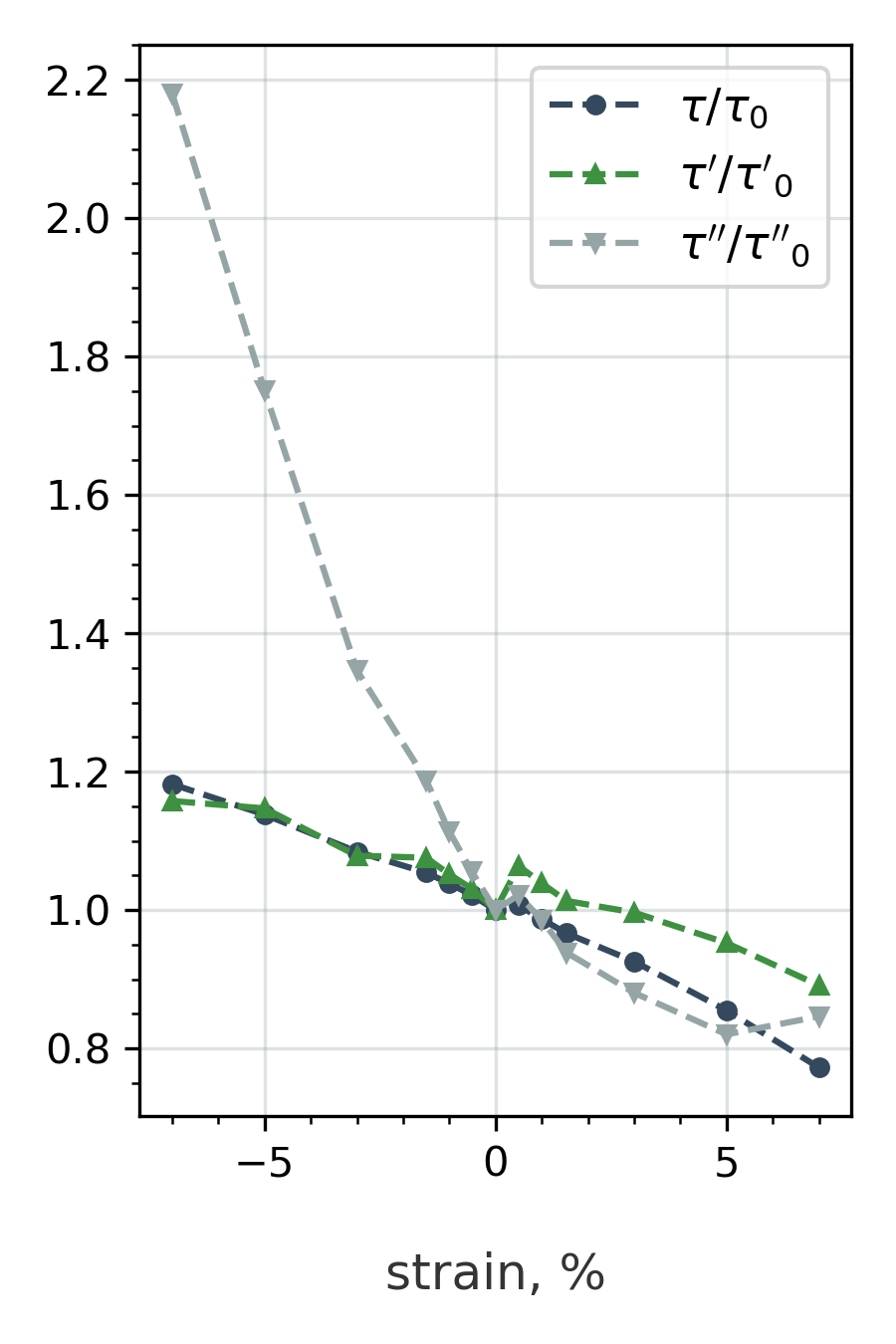}\label{tau_normalized}
	\llap{\parbox[b]{2.5in}{\textbf{b}\\\rule{0ex}{3.40in}}}
	\caption{\textbf{Dependency of $J_\mathrm{eff}=1/2$ overlap on the strain.} \textbf{a} NN, 2NN and 3NN hopping parameters ($\tau$, $\tau'$ and
		$\tau''$, correspondingly) between $J_\mathrm{eff}=1/2$ states
		calculated as a superposition of obtained from DFT calculations
		$xy$, $yz$ and $xz$ hopping parameters. \textbf{b} Relative change of 
		NN, 2NN and 3NN $J_\mathrm{eff}=1/2$ overlap normalized by their values in
		pristine condition ($\tau_0$, $\tau_0'$ and $\tau_0''$, correspondingly.
		2NN and 3NN hoppings for $t_{\mathrm{2g}}$ orbitals used to calculate $\tau$'s are
		shown in SI.)
		\label{tau_both}}
\end{figure}

We now estimate the overlap between the $J_\mathrm{eff}=1/2$ states for NN, 2NN, and 3NN (denoted $\tau$, $\tau'$ and
$\tau''$, correspondingly), which can be calculated from $t_{\mathrm{2g}}$
orbitals overlap using the Clebsh-Gordon coefficients.~\cite{Jackeli2009, Plotnikova2016}
In Fig.~\ref{tau_both}, we show the change of the overlap of NN $J_\mathrm{eff}=1/2$
states calculated from the DFT values obtained here.
As experimental values of $J_\mathrm{eff}=1/2$
states overlap $\tau$ are hard to measure, one can try to compare hopping parameters $\tau$ with available experimental
estimates of magnetic exchange interactions $\mathcal{J}$ (see Table 1 in Ref.~\onlinecite{Paris2020}), which in ambient conditions are assumed 
to scale with $\tau^2$.~\cite{Jackeli2009} However, in a recent RIXS study on strained Sr$_2$IrO$_4$,~\cite{Paris2020},
the authors suggest that the simple $\mathcal{J} \propto \tau^2/U$ relationship
fails for strained {\sio} due to the polaronic renormalization of the
charge excitations. In particular,
the first-neighbor exchange interaction $\mathcal{J}_\mathrm{1}$ was shown to decrease slightly
upon the tensile strain, while $\mathcal{J}_\mathrm{2}$ and $\mathcal{J}_\mathrm{3}$ decreased much
faster, based on a fit of the Heisenberg model to the measured
magnon dispersion.~\cite{Paris2020} 
An earlier RIXS study also suggested that magnetic exchange
interaction $\mathcal{J}$ increases upon the compressive strain.~\cite{LupascuPRL2014}

%%%%%%%%%%%%%%%% FIG 5 %%%%%%%%%%%%%
%%%%%%%%%%%%%%%% FIG 5 %%%%%%%%%%%%%

Consistent with both RIXS
studies,~\cite{LupascuPRL2014,Paris2020} calculated here values of NN, 2NN, and 3NN hopping parameters are all
decreasing upon tensile strain (Fig.~\ref{tau_both}). As discussed in Ref.~[\onlinecite{Paris2020}], this trend for $\tau$'s is
significantly slower than that observed for superexchange interaction $\mathcal{J}$, indicating that $\mathcal{J} \propto \tau^2/U$ relationship indeed fails for strained {\sio}. Our DFT strain trends are also consistent with the 
modest increase of magnetic exchange interaction $\mathcal{J}_\mathrm{1}$ under
compressive strain reported in the two-magnon Raman study.~\cite{SeoPRB2019}

It is interesting to compare the trends observed in {\sio} to those in
$3d$ transition metal oxides - cuprates. In
Ref.~[\onlinecite{IvashkoNatCom2019}] authors used XAS at Cu $L_3$-edge of
La$_2$CuO$_4$ together with analytical and DFT theoretical approaches to
show that both bandwidth and electron-electron correlations were
increasing upon the compressive strain. As a result, magnetic exchange
$\mathcal{J}$ was
shown to increase (decrease) almost linearly upon the compressive (tensile)
strain. In comparison, while the orbital-dependent hoppings in {\sio} behave very
differently from this (Fig.~\ref{t_and_theta_d_under_strain}b),
showing the surprising decrease in $t_1$ with compressive strain,
effective $J_\mathrm{eff}=1/2$ orbitals in {\sio} have strain
dependence (Fig.~\ref{tau_both}) somewhat similar to those of
cuprates, echoing the famous parallel between {\sio} and
La$_2$CuO$_4$.

\subsection{The evolution of the Fermi surface under strain.}

Structural response to strain may also be accompanied by changes in the Fermi surface. Thus, strong changes of Fermi surface upon uniaxial pressure have recently been reported in Ru-based compound Sr$_2$RuO$_4$,~\cite{SunkoNPJQuantMat2019,BarberPRB2019} along with the more than double increase of the superconducting transition temperature.~\cite{TaniguchiJPhysSocJap2015}
A recent work on Sr$_2$IrO$_4$ employing a tight-binding model has pointed out that out-of-plane tilting of the oxygen octahedra can induce shrinking of the Fermi surface and suppress nesting and the $d$-wave superconductivity.~\cite{LindquistPRB2019} There is no out-of-plane tilting in {\sio} under pristine conditions or modest strain - it was very recently shown to appear only under the pressure of as much as 40 GPa.~\cite{ChunhuaChen2019} However, in-plane octahedra rotation is strongly affected by the modest strain already, and it is important to understand if and how the Fermi surface is affected, particularly on the beyond-mean-field level.

%%%%%%%%%%%%%%%% FIG 6 %%%%%%%%%%%%%
\begin{figure}
	\centering
	\includegraphics[width=0.48\linewidth,trim= 0 100 0 0, clip]{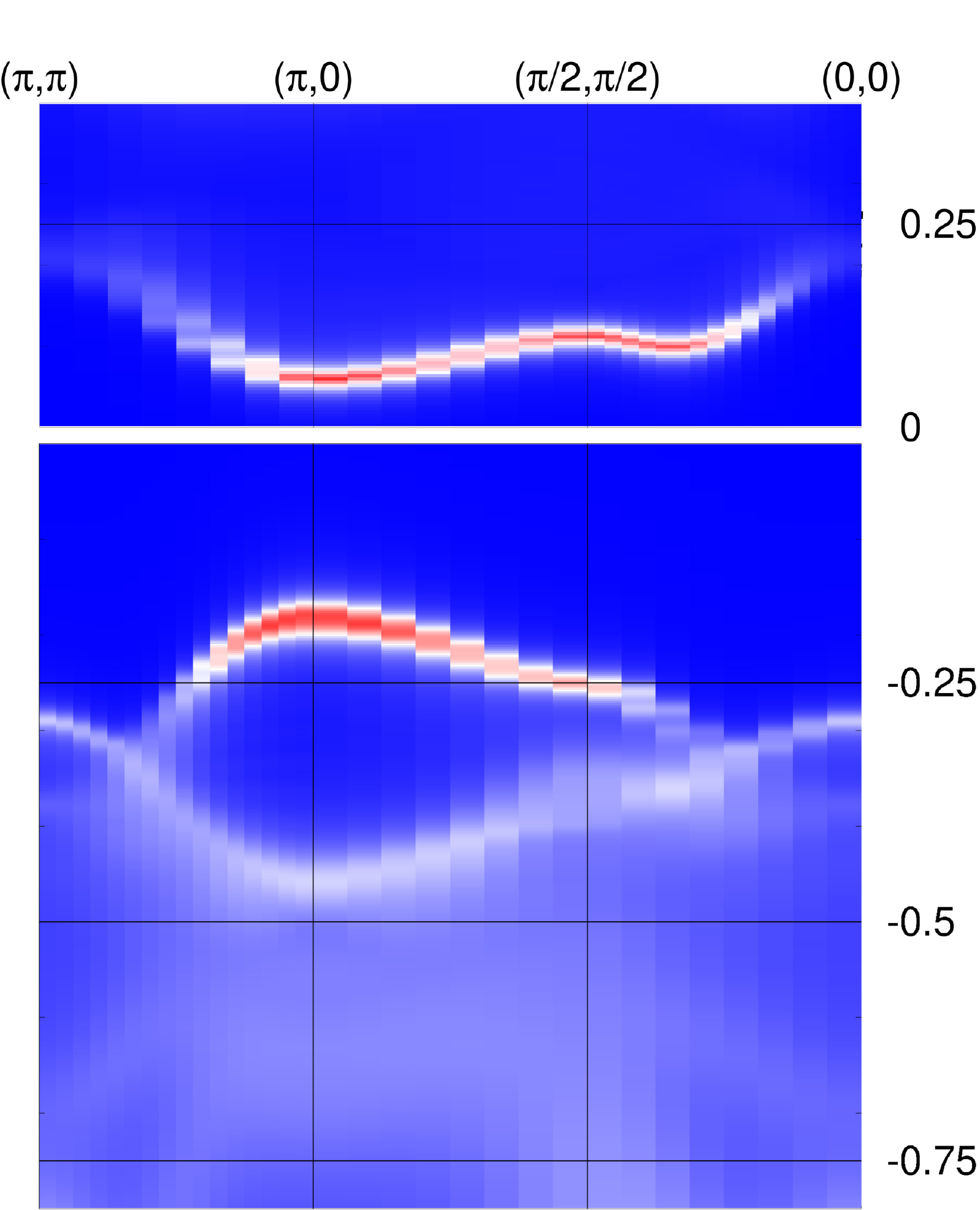}\label{PES_LSAT}
	\llap{\parbox[b]{2.95in}{\textcolor{white}{\textbf{a)} -0.52\% \\\rule{0ex}{2.60in}}}}
	\includegraphics[width=0.48\linewidth,trim= 0 100 0 0, clip]{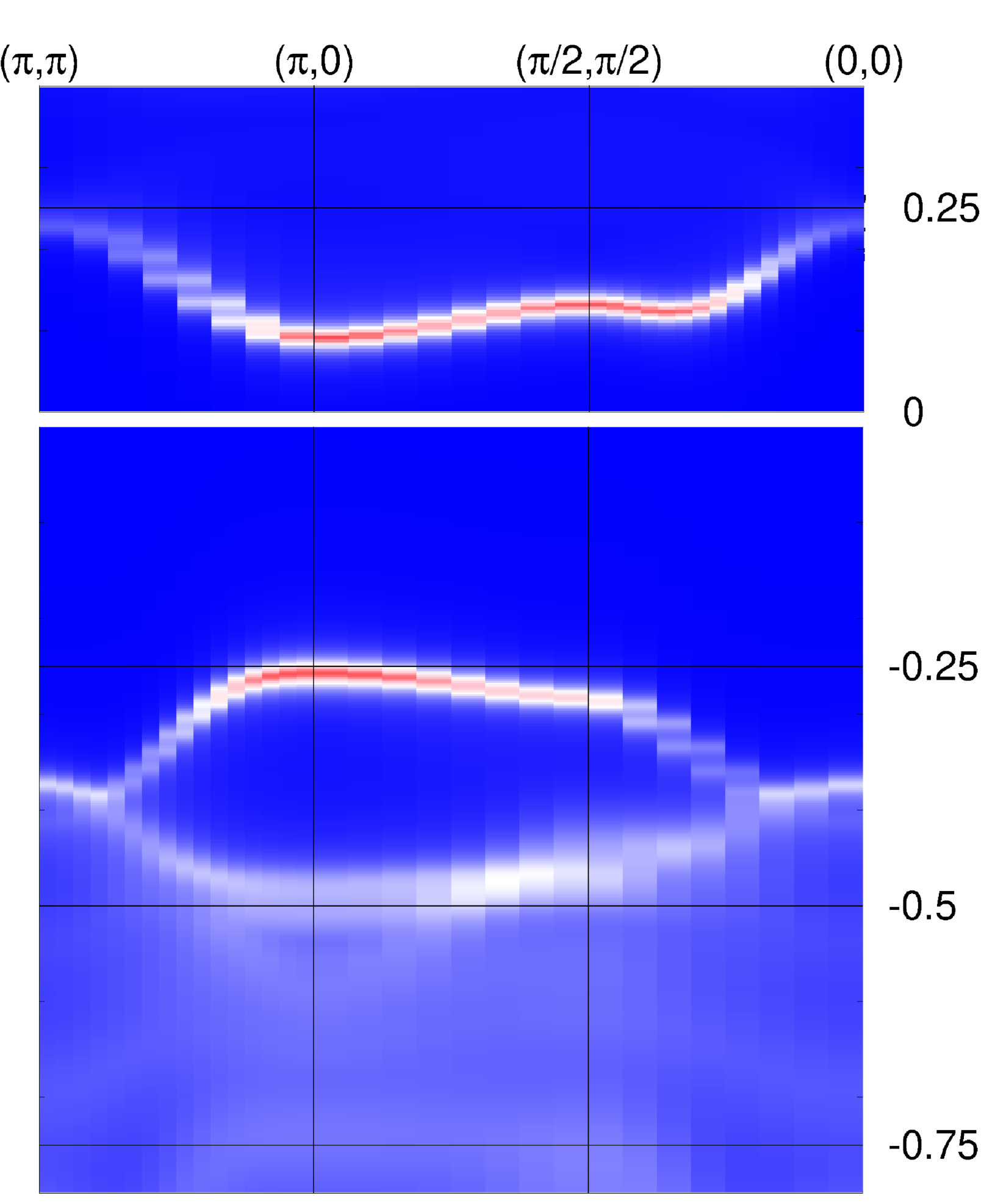}\label{PES_GSO} %trim=left bottom right top, clip
	\llap{\parbox[b]{2.95in}{\textcolor{white}{\textbf{b)} +1.53\% \\\rule{0ex}{2.60in}}}}
	\caption{\textbf{Dependency of the spectral function of {\sio} on the strain.} Spectral function calculated for various strain values: \textbf{a} LSAT, -0.52\% (compressive strain), \textbf{b} GSO, +1.53\% (tensile strain) corresponding to ARPES (negative energies) and inversed photoemission spectra (positive energy). The horizontal axis is the 2D crystal momentum. The vertical axis is the energy (eV), where zero energy represents the Fermi level. 
		\label{PES_IPES}}
\end{figure}
%%%%%%%%%%%%%%%% FIG 6 %%%%%%%%%%%%%

To study the evolution of the Fermi surface under strain we
calculate photoemission spectral functions of strained Sr$_2$IrO$_4$,
using extended $t-{\mathcal{J}}$ model formalism developed in Ref.~[\onlinecite{Paerschke2017}]. 
The extended $t-{\mathcal{J}}$ model used in the
calculation (see Methods) depends on two sets of parameters: the
magnetic exchange parameters $\mathcal{J}_\mathrm{1}$, $\mathcal{J}_\mathrm{2}$,
$\mathcal{J}_\mathrm{3}$, Ising anisotropy coefficient $\Delta$,\cite{ValePRB2015}, and the hopping parameters $t_{i}$ describing overlap of the $t_{\mathrm{2g}}$ orbitals. We obtain the set of $t_{i}$'s for each strain
value from DFT calculations as discussed in detail above. Using this Wannier Hamiltonian as a starting point describing single-particle hopping processes, we consider all possible many-body hopping processes to derive the hopping part of the extended $t-{\mathcal{J}}$ model.~\cite{Paerschke2017}

To properly account for the changes in the electronic structure, we need to account for the evolution of the 
magnetic exchange parameters with the strain. It has been obtained from the published\cite{Paris2020} fits to the RIXS spectra on strained samples. As experimental data is available for small strain values range only, we restrict calculations of
photoemission spectra to that range and show the photoemission (and inverse photoemission) spectra in Fig.~\ref{PES_IPES} for two substrates: (LaAlO${_3}$)${_0.3}$(Sr$_2$AlTaO$_6$)${_0.7}$ (100) (LSAT) and GdScO$_3$ (110) (GSO), providing a strain of -0.52$\%$ and +1.53$\%$,
correspondingly.

%%%%%%%%%%%%%%%% FIG 7 %%%%%%%%%%%%%
\begin{figure}
	\centering
	\includegraphics[width=0.48\linewidth,trim= 0 0 0 0, clip]{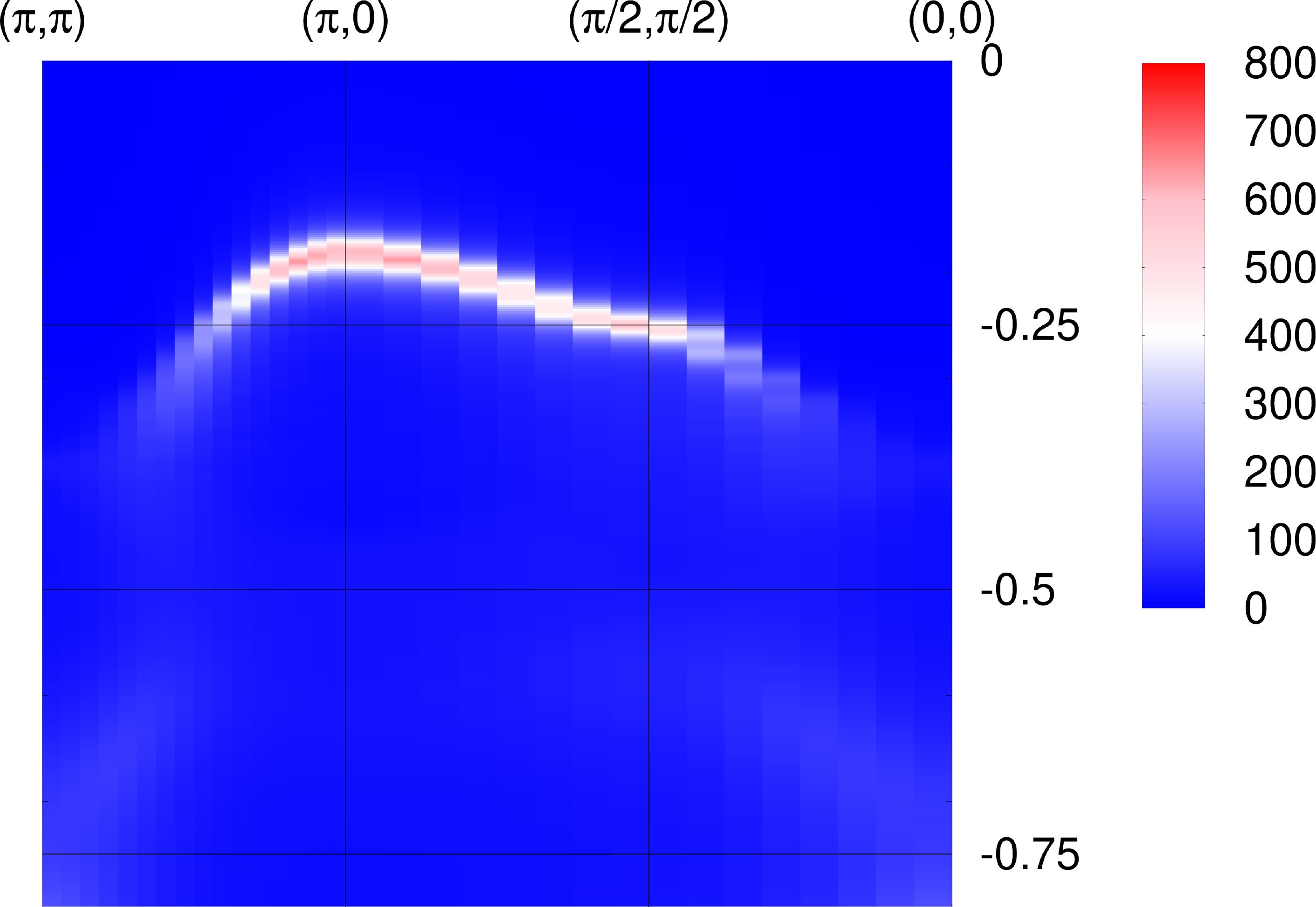}
	\label{PES_LSATJ0}
	\llap{\parbox[b]{2.95in}{\textcolor{white}{\textbf{a} -0.52\%, $J=0$ \\\rule{0ex}{1.75in}}}}
	\includegraphics[width=0.48\linewidth,trim= 0 0 0 0, clip]{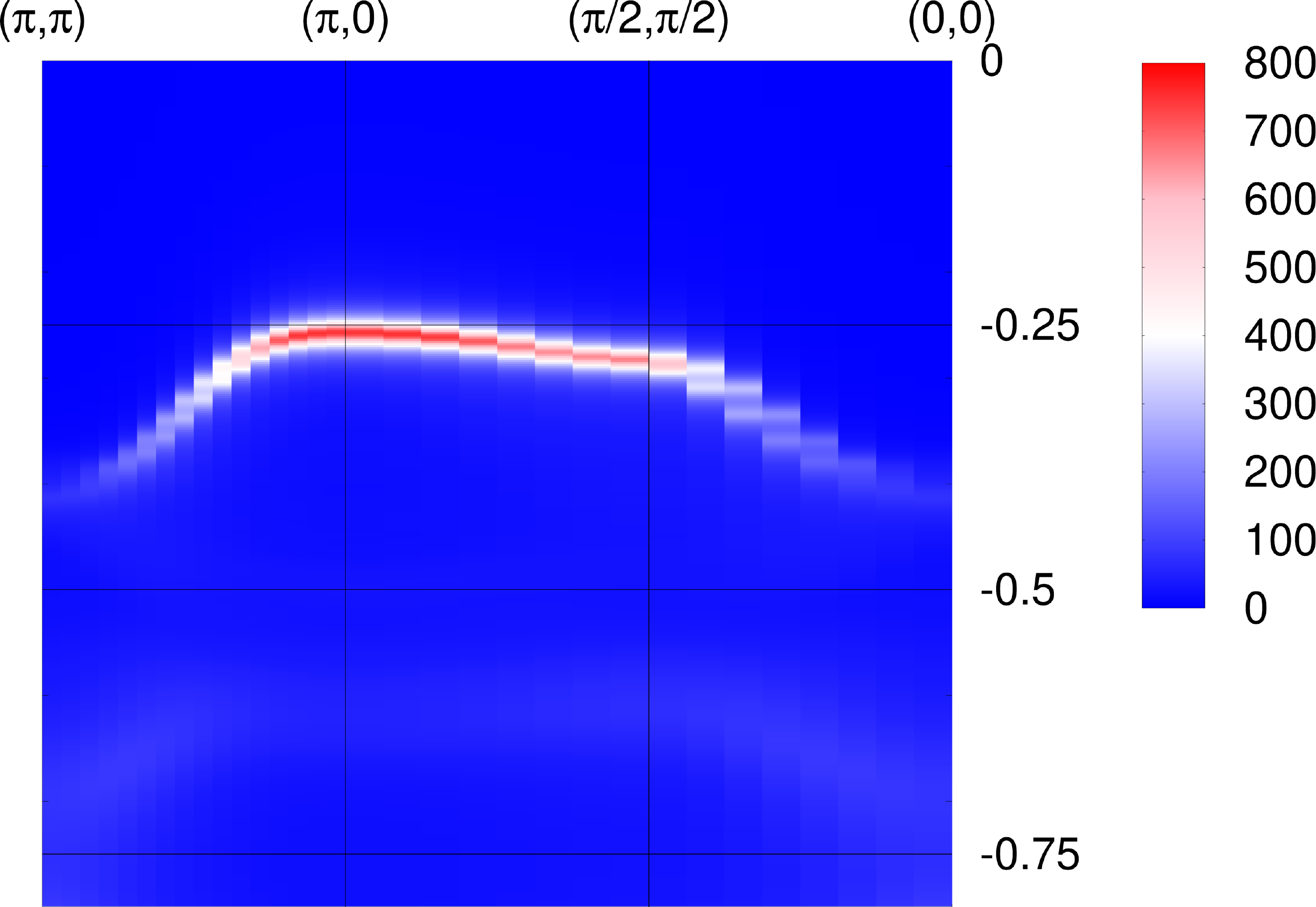}
	\label{PES_GSOJ0}
	\llap{\parbox[b]{2.95in}{\textcolor{white}{\textbf{b} +1.53\%, $J=0$ \\\rule{0ex}{1.75in}}}}
	\includegraphics[width=0.48\linewidth,trim= 0 0 0 0, clip]{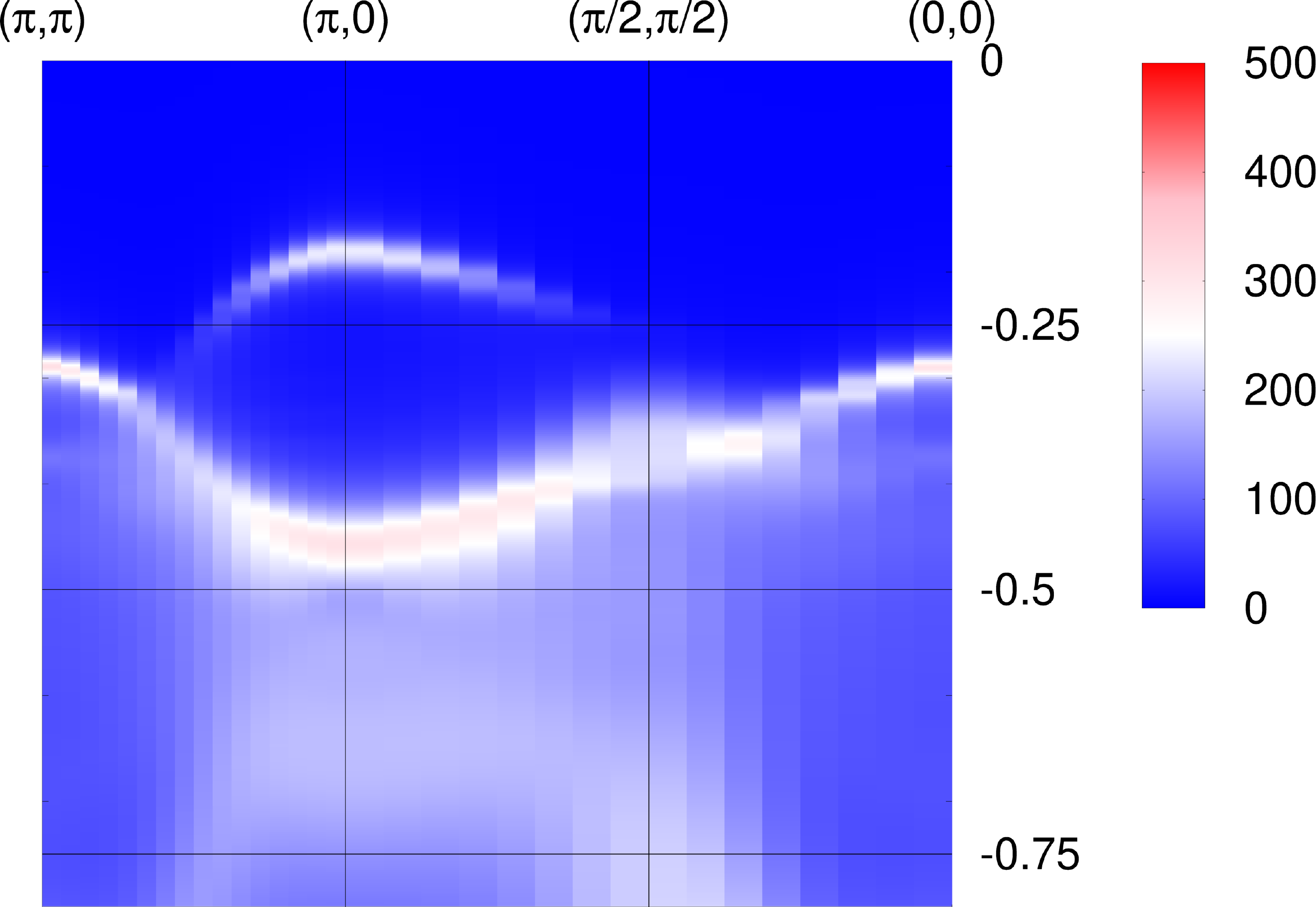}
	\label{PES_LSATJ1} %trim=left bottom right top, clip
	\llap{\parbox[b]{2.95in}{\textcolor{white}{\textbf{c} -0.52\%, $J=1$ \\\rule{0ex}{1.75in}}}}
	\includegraphics[width=0.48\linewidth,trim= 0 0 0 0, clip]{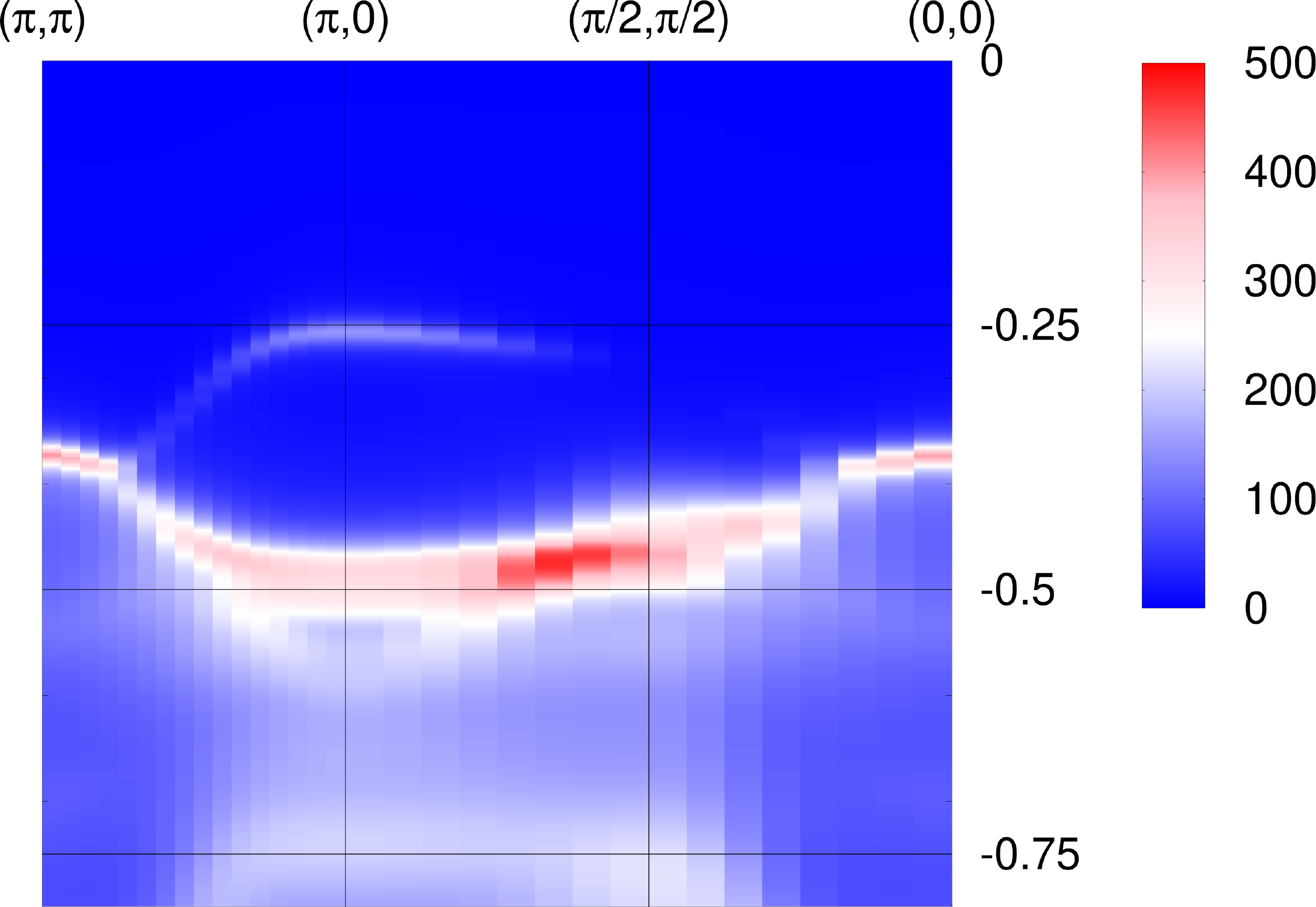}
	\label{PES_GSOJ1} 
	\llap{\parbox[b]{2.95in}{\textcolor{white}{\textbf{d} +1.53\%, $J=1$ \\\rule{0ex}{1.75in}}}}
	\caption{\textbf{Contribution of singlet $J=0$ and triplet $J=1$ states to the spectral function of {\sio} for various strain values.} \textbf{a}-\textbf{b}: singlet $J=0$ contribution to the ARPES spectral function: \textbf{a} LSAT, -0.52\% (compressive strain), \textbf{b} GSO, +1.53\% (tensile strain); \textbf{c}-\textbf{d}: triplet $J=1$ contribution to the ARPES spectral function: \textbf{c} LSAT, -0.52\% (compressive strain), \textbf{d} GSO, +1.53\% (tensile strain). The horizontal axis is the 2D crystal momentum. The vertical axis is the energy (eV), where zero energy represents the Fermi level.
		\label{PES_J1_J0}}
\end{figure}
%%%%%%%%%%%%%%%% FIG 7 %%%%%%%%%%%%%

Calculated photoemission spectra show one conductance band at positive energies, and two valence bands at negative energies: a sharp singlet band around -0.25 eV and a more incoherent triplet band at -0.5 eV. (see Fig.~\ref{PES_J1_J0} for a detailed discussion on the band character).
We see that the strain-induced changes of the photoemission spectra are quite prominent for samples with a strain difference of $2 \%$.
First, for tensile strain, as compared to compressive strain, % there are few noteworthy features: First%, both bands shift and 
the Mott gap increases (Fig.~\ref{PES_IPES}b), suggesting stronger polaron binding of
the photoinduced hole to the magnetic background.~\cite{Paris2020}
Second, upon compressive strain, the photoemission spectra of {\sio} show a highly dispersive singlet band (Fig.~\ref{PES_IPES}a), while upon tensile strain, both
singlet and triplet bands are much less dispersive, %the dispersion of the valence band changes upon tensile strain significantly: 
and the Fermi surface of \sio becomes relatively flat
(Fig.~\ref{PES_IPES}b). It is important to note that the relative flattening of
the Fermi sheet upon tensile strain is a many-body effect distinct
from the anisotropic compass-like hoppings under compressive strain.
One should be able to observe such significant renormalization of
the spectral weight in the ARPES data even for small, realistic values of 
strain.

The conduction band is only weakly affected by strain. Fig.~\ref{PES_IPES} shows a marginal flattening of the conduction band upon the tensile
strain. We, therefore, expect a minimal
effect of epitaxial strain on possible superconductivity. 

To explore the effect of strain on the ARPES spectra in more detail, we
plot in Fig.~\ref{PES_J1_J0} separate contributions of the singlet
($J=0$) and triplet ($J=1$) charge excitations to the full spectra. As
one can see, under tensile strain, $J=0$ contributes the most at
($\mathrm{\pi}$,0), whereas under compressive strain, its contribution is
slightly more  widespread. On the contrary, the $J=1$ spectral weight
is shifted to $(\mathrm{\pi}/2,\mathrm{\pi}/2)$ at the tensile strain and to $(\mathrm{\pi},0)$ at the
compressive strain. In particular, $J=1$ contribution to the "lower
energy" band of the photoemission spectra is strongly reduced upon the
tensile strain. We thus observe strain-controlled spectral weight
redistribution between the charge carriers of singlet and triplet
characters. Already moderate tensile strain
is sufficient to make the lower energy band of almost purely singlet
character.

In summary, we predict a dramatic strain dependence of the electronic
properties of {\sio} for compressive v/s tensile strain. 
The most remarkable feature is the appearance of the compass-model-like contribution of electron propagation due to the separation of the Fermi sea in {\sio} 
into two subsets of bond- and orbital-dependent carriers
under compressive strain. This enables the formation of charge density wave due to nesting and could be connected to the
 puzzling metalization avoidance in
{\sio} upon pressure. The Fermi sea separation originates from strain dependency of relative Ir-O hybridization, as well as on-site O energy, suggesting an important role of oxygens in a low-energy physics of strained iridates and other transitional-metal oxides.

Despite the suppression of the bond-independent $xy$ hopping $t_1$ 
under compressive strain, the hopping amplitude of the composite $J_\mathrm{eff}=1/2$
state still increases under compressive strain owing to the contribution from hoppings between
$xz/yz$ orbitals ($t_2$) --- the
largest of two in-plane nearest-neighbor direction-dependent
hoppings. The obtained trends for $J_\mathrm{eff}=1/2$ hopping, $\tau$,
are in good agreement with available experimental data.

We also calculated the photoemission spectra of {\sio} upon compressive and
tensile strain (for samples grown on LSAT and GSO substrates,
respectively). We
find that under compressive (tensile) strain,
the singlet band becomes significantly more (less) dispersive, and both
the singlet and triplet bands shift up (down) in energy. We also show
that the electronic properties of the low-energy model can be controlled 
by strain, since the already moderate tensile strain is sufficient to make
the lower energy band of almost purely singlet character, and shift the
triplet spectral weight to $(\mathrm{\pi}/2,\mathrm{\pi}/2)$ point. These features can be
readily observed in the future ARPES measurements --- a smoking gun test
of our findings.

\begin{methods}

\subsection{DFT calculations.}

The DFT results are calculated by the Quantum Espresso package
~\cite{Giannozzi_2009} using Perdew-Burke-Ernzerhof (PBE)
      exchange-correlation functional~\cite{PerdewPRL1996} and
projector-augmented-wave pseudopotentials~\cite{BloechPRB1994} with
the unit cell containing 16 Sr, 8 Ir, and 32 O atoms. In the calculations
of structure relaxation and electronic structure, spin-orbit interaction
is not included. The unstrained structure (a = 5.5098 \AA, c = 26.1522
\AA) is acquired through relaxing both the lattice geometry and the
atomic positions. For biaxial-strained structures, the $a$- and $b$-axis are
fixed as per the strain value while the $c$-axis and the atomic
positions are simultaneously relaxed. The plane
wave cutoff energy is set to be 40 Ry, and a $\Gamma$-centered $10
\times 10 \times 2$ $k$-points in the full Brillouin zone are sampled. The
structural and electronic convergence criteria were set to be $10^{-4}$
Ry/$a_0$ and $10^{-6}$ Ry, respectively.

With the DFT eigenvalues and eigenvectors, we then use Wannier90 package~\cite{MOSTOFI2008685} 
and implement a disentanglement procedure to obtain
the Wannier functions and the hopping parameters. During the
disentanglement, 120 bands were
considered, which accounted for contributions from 96 (32$\times$3)
O-$2p$
orbitals and 24 (8$\times$3) Ir-$t_{\mathrm{2g}}$ orbitals. Four iridium layers (i.e. 8 Ir atoms) are involved to construct 24 wannierized $t_{\mathrm{2g}}$ orbitals in one unit cell (6 Wannier functions per Ir layer).

\subsection{Calculation of the photoemission spectra.}
The ARPES spectral function is obtained as imaginary part of the Green function $G ({\bf k},\omega)$ describing propagation of the photohole in the ground state, dressed in the low-energy magnetic excitations (magnons): 
$G ({\bf k},\omega)={\rm Tr} {\langle {\rm AF} | { \bf h }_{ {\bf
    k}}\frac{1}{\omega-\mathcal{H}_{\rm t-{\mathcal{J}}}+i\delta}
	{\bf h}_{ {\bf k}}^\dagger|  {\rm AF} \rangle}$. Here, photohole ${ \bf h }_{ {\bf k}}$ is a vector in a full spin-orbital Hilbert space, obtained by projecting three $t_{\mathrm{2g}}$ orbitals of Ir atom onto the spin-orbit coupled basis~\cite{Paerschke2017,PaerschkeRay2018}. 
    The motion of the hole is governed by $t-{\mathcal{J}}$ Hamiltonian
$\mathcal{H}_{\rm t-{\mathcal{J}}}=\mathcal{H}_{\rm t}+\mathcal{H}_{\mathcal{J}}$.  The
magnetic $H_{\mathcal{J}}$ part includes NN, 2NN and 3NN Heisenberg
interactions $\mathcal{J}_\mathrm{1}$ , $\mathcal{J}_\mathrm{2}$, $\mathcal{J}_\mathrm{3}$ and an anisotropic $\Delta$ term, arising from non-negligible Hund's coupling~\cite{ValePRB2015,PinciniPRB2017}:
\begin{eqnarray}
&\mathcal{H}_{\mathcal{J}}=\mathcal{J}_\mathrm{1}\sum_{\langle i,j \rangle}{\left(\mathbf{\tilde{S}}_i^x\mathbf{\tilde{S}}_j^x+\mathbf{\tilde{S}}_i^y\mathbf{\tilde{S}}_j^y+(1-\Delta)\mathbf{\tilde{S}}_i^z\mathbf{\tilde{S}}_j^z\right)}\\
    & +\mathcal{J}_\mathrm{2}\sum_{\langle\langle i,j \rangle\rangle}{\mathbf{\tilde{S}}_i\cdot\mathbf{\tilde{S}}_j}
    +\mathcal{J}_\mathrm{3}\sum_{\langle\langle\langle i,j \rangle\rangle\rangle}{\mathbf{\tilde{S}}_i\cdot\mathbf{\tilde{S}}_j} \nonumber
\end{eqnarray}
The hopping $\mathcal{H}_{\rm t}$ part is derived by projecting multiorbital Hubbard model employing orbital-dependent hopping parameters $t_{i}$ onto spin-orbit coupled basis.~\cite{Paerschke2017,PaerschkeRay2018} The motion of a charge excitation in the new spin-orbit coupled basis is then expressed analytically in terms of these $t_{i}$ which are obtained directly from DFT.~\cite{Paerschke2017,PaerschkeRay2018} We evaluate the Green function $G ({\bf k},\omega)$ using
the self-consistent Born approximation (SCBA).\cite{Martinez1991} SCBA is a diagrammatic approach 
that evaluates Green function of a quasiparticle dressed with bosons (here, photohole dressed with magnons) that form diagrams of
rainbow type.~\cite{Martinez1991} The spectral functions are calculated
numerically for a $16 \times 16$ cluster.

\end{methods}

\begin{addendum}
\item[Data availability]
All the data that support the findings of this study are available from the
corresponding author (E.M.P.) upon reasonable request.

\item[Code availability]
The codes used in this study are available from the corresponding author (E.M.P.) on
reasonable request.

\item E.M.P. thanks Eugenio Paris, Thorsten Schmitt, Krzysztof Wohlfeld and other coauthors for
an inspiring previous collaboration,~\cite{Paris2020} and is grateful to
Gang Cao, Ambrose Seo and Jungho Kim for insightful discussions. R.R.
acknowledges helpful discussion with Sanjeev Kumar %(IISER Mohali, India) 
and Manuel Richter. %(IFW Dresden, Germany). 
This project has received funding from the European Union's Horizon 2020 research and innovation
programme under the Marie Sklodowska-Curie grant agreement No 754411. C.C.C. acknowledges support from the U.S. National Science Foundation Award No. DMR-2142801.
%%%%%%%%%%%%%%%%%%%%%%%%%%%%%%%%%%%%%%%%%%%%%%%%%%%%
%%%%%%%%%%%%%%%%%%%%%%%%%%%%%%%%%%%%%%%%%%%%%%%%%%%%

\item[Competing Interests] The authors declare that there are no competing interests.

\item[Author Contributions] 

E.M.P. designed the project and carried out the photoemission spectra calculations. E.M.P. and R.R.
performed the Slater-Koster analysis. W.C.C. carried out the DFT and Wannier function analysis. E.M.P.
and C.C.C. supervised the project. All the authors analyzed the results. E.M.P. wrote the manuscript with the help of R.R. and C.C.C..

\item[Correspondence] Correspondence and requests for materials should
    be addressed to E. M. P{\"a}rschke ~(email:
    ekaterina.paerschke@ist.ac.at).
\end{addendum}

      \clearpage
      \newpage
      %\documentclass{nature}
%
%\usepackage{amsmath,amssymb}
%\usepackage{epsfig}
%\usepackage{subfigure}
%\usepackage{graphicx}
%\usepackage{todonotes}
%\usepackage{dcolumn}
%\usepackage{bm}
%\usepackage[colorlinks,linkcolor=blue,hyperindex,CJKbookmarks]{hyperref}
%\usepackage{float}
%\usepackage{hyperref}
%\usepackage{comment}
%\usepackage{color}
%\usepackage{lineno}
%%\linenumbers
%\hyphenpenalty=1000
%\tolerance=2000
%
%%added by rr to strike-out text
%\usepackage{soul}
%
%%equivalent of onlinecite in APS
%\newcommand{\onlinecite}[1]{\hspace{-1 ex} \nocite{#1}\citenum{#1}}
%
%
%\newcommand{\kp}[1]{\textcolor{black}{#1}}
%\newcommand{\md}[1]{\textcolor{black}{#1}}
%\newcommand{\change}[1]{\textcolor{red}{#1}}
%\newcommand{\rr}[1]{\textcolor{magenta}{#1}}
%\newcommand{\remark}[1]{\textcolor{black}{[{\bf RR:} #1]\todo{check
%this out}}}
%
%\newcommand{\tinytodo}[2][]
%{\todo[caption={#2}, size=\footnotesize, #1]{\renewcommand{\baselinestretch}{0.5}\selectfont#2\par}}
%
%\newcommand{\sio}{{Sr$_2$IrO$_4$}}
%
%
%%\title{Supplemental Information: \newline Evolution of electronic and magnetic properties of Sr$_2$IrO$_4$ under strain}
%\title{Supplemental Information}
%
%\author{Ekaterina M. P{\"a}rschke$^{1,	*}$, Wei-Chih Chen$^{2}$,
%Rajyavardhan Ray$^{3,4,5}$ \& Cheng-Chien Chen$^{2}$}
%
%
%%to show graphics
%\makeatletter
%\let\saved@includegraphics\includegraphics
%\AtBeginDocument{\let\includegraphics\saved@includegraphics}
%\renewenvironment*{figure}{\@float{figure}}{\end@float}
%\makeatother
%
%\renewcommand{\baselinestretch}{1.50} 
%%\setlength{\baselineskip}{0pt} 
%
%for relabeled figures, tables, etc.
\renewcommand{\thepage}{S\arabic{page}}
\renewcommand\tablename{Supplementary Table}
\renewcommand\figurename{Supplementary Figure}
\setcounter{figure}{0}
\setcounter{page}{1}

\section*{\Large{\textsf{Supplemental Information}}}
%\section*{Supplemental Information}
%\section*{Supplementary Note 1}
\subsection{}
We start by presenting hopping parameters for further neighbors
extracted from DFT. In Supplemental Figure~\ref{t_2NN_3NN_4NN_nondiag_under_strain}, we show the hopping parameters for further neighbors: second nearest (2NN) and third nearest (3NN) neighbors. These values were extracted from DFT calculations using Wannier functions fitting, as discussed in detail in Methods.

%%%%%%%%%%%%%%%% FIG Ap 1%%%%%%%%%%%%%
\begin{figure}[!h]
	\centering
	\includegraphics[width=0.4\linewidth]{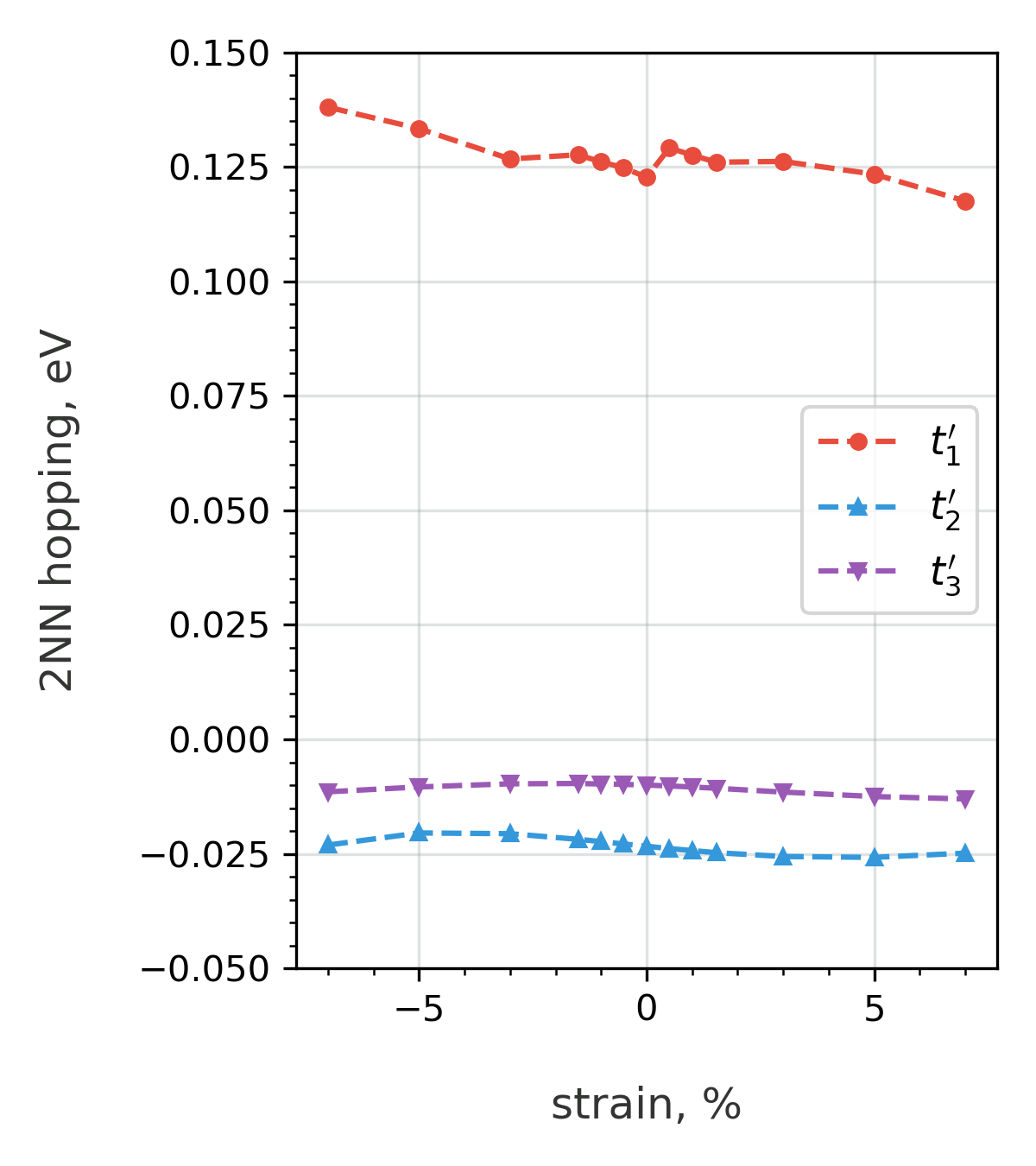}\label{t2NNDFT}
	\llap{\parbox[b]{2.50in}{\textbf{a}\\\rule{0ex}{2.80in}}}
	\includegraphics[width=0.4\linewidth]{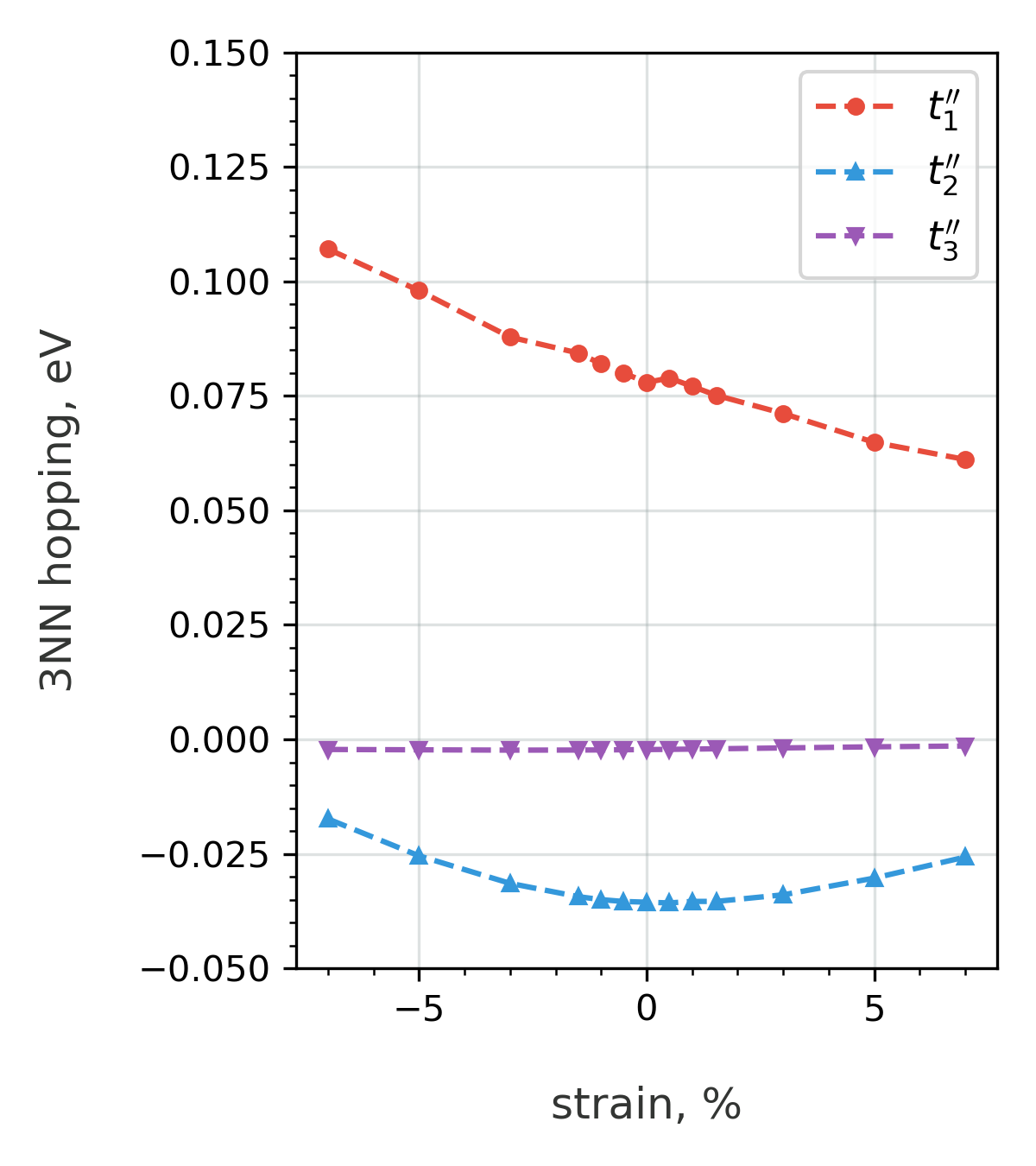}\label{t3NNDFT}
	\llap{\parbox[b]{2.50in}{\textbf{b}\\\rule{0ex}{2.80in}}}
	\caption{
		(a) 2NN hopping parameters, (b) 3NN hopping parameters as a function of strain.
		Hopping parameters were extracted from the Wannier function fit of the DFT band structure. Dashed lines are used to show Wannier function fitting results, while solid lines are used to show results obtained using Slater-Koster integrals.
		\label{t_2NN_3NN_4NN_nondiag_under_strain}
	}
\end{figure}
%%%%%%%%%%%%%%%% FIG Ap 1%%%%%%%%%%%%%

%\section*{Supplementary Note 2}
\subsection{}
\begin{figure}[!h]
	\centering
	\includegraphics[width=0.3525\linewidth]{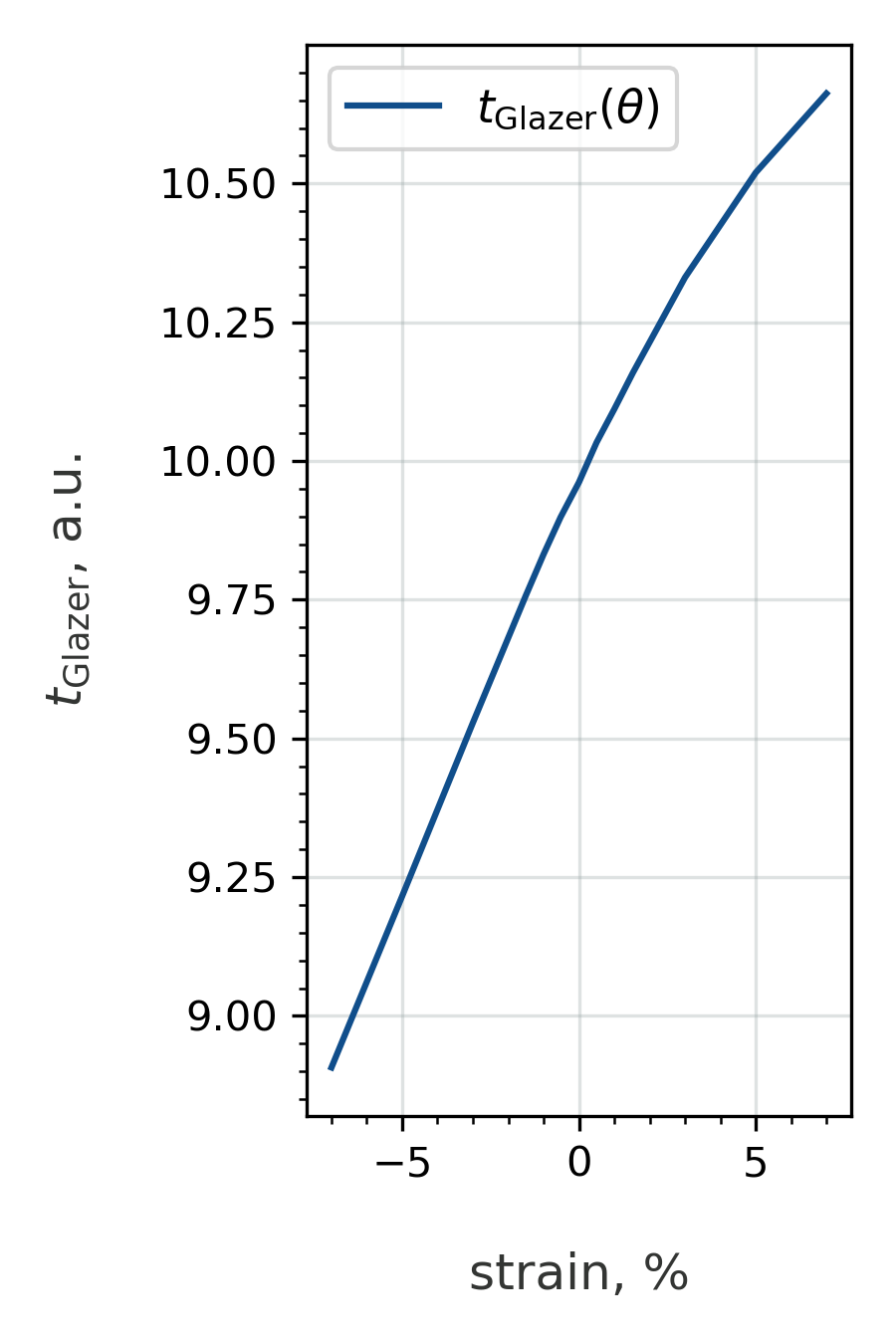}\label{tGlazertheta}
	\llap{\parbox[b]{2.10in}{\textbf{a}\\\rule{0ex}{3.3in}}}
	\includegraphics[width=0.29375\linewidth]{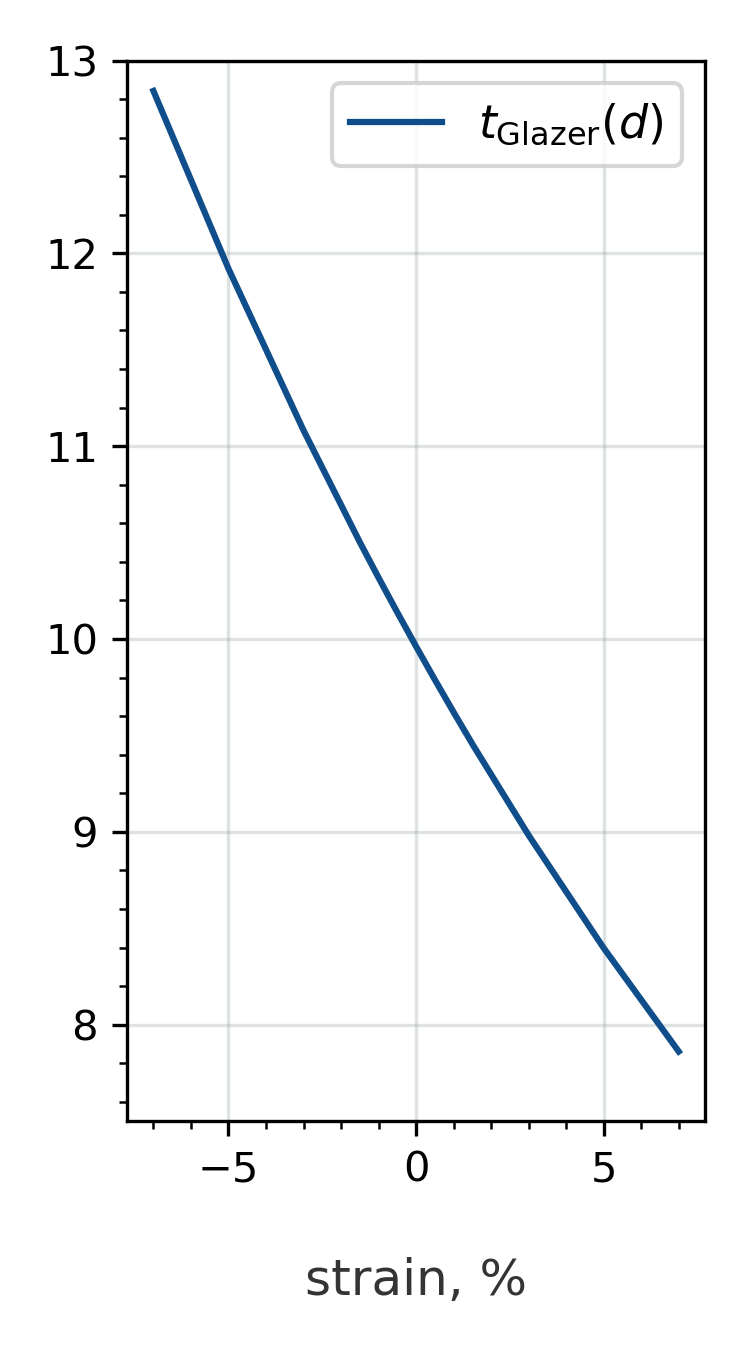}\label{tGlazerd}
	\llap{\parbox[b]{2.05in}{\textbf{b}\\\rule{0ex}{3.3in}}}
	\includegraphics[width=0.29375\linewidth]{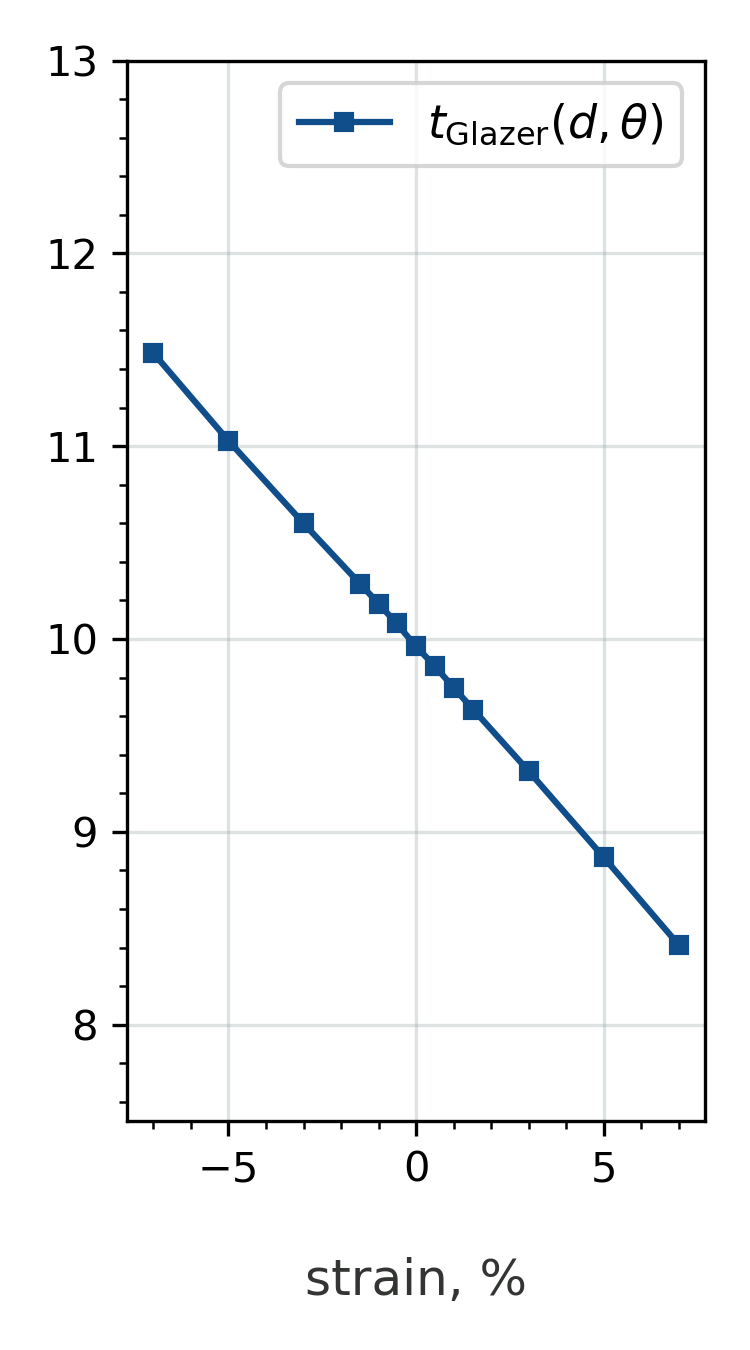}\label{tGlazerthetad}
	\llap{\parbox[b]{2.05in}{\textbf{c}\\\rule{0ex}{3.3in}}}
	\caption{ (a)-(c) Value of phenomenological hopping parameter $t_{\rm{Glazer}}$ as a function of (a) octahedra in-plane rotation angle $\theta$, (b) Ir-Ir distance $d$, (c) both $\theta$ and $d$ (calculated using values of $\theta$ and $d$ extracted from DFT). 
		\label{Glazer}}
\end{figure}

We proceed with Glazer's picture of the rigid octahedral rotation that is often used for
perovskite metal oxides. 
%Consequently, 
In this approach, %the distance $d$ is assumed to change slowly while 
the main effect of the strain is assumed to be the
change of the in-plane rotation angle $\theta$ \cite{Glazer1972}. 
%\kp{We thus neglect} the change of bond length $d$ and only take into account the change of the angle %\remark{the first part of the sentence seems repetitive} 
Using strain-dependent angle values extracted from our DFT calculations,
we get that the overall hopping $t \propto \cos^{4.5}\theta/(0.5 d)^{3.5}$
should decrease under compressive strain (Supplemental Figure \ref{Glazer}(a)), as indeed
expected for metal oxides with perovskite structure\cite{SeoPRB2019}. This trend, however, clearly contradicts those obtained in DFT calculations in the main text. Even when changes in the bond length $d$ are taken into
account (Supplemental Fig \ref{Glazer}(b)-(c)), this simple phenomenological approach could only explain the evolution of direction-dependent $t_2$ and $t_3$ hoping parameters, but not the $xy-xy$ hopping parameter
$t_1$. It is, moreover, not consistent with the two-magnon Raman
scattering data, as pointed out by [\onlinecite{SeoPRB2019}]. We thus conclude that simple Glazer's rigid octahedra picture is not applicable for iridates.

%%%%%%%%%%%%%%%% FIG 3 %%%%%%%%%%%%%
%%%%%%%%%%%%%%%% FIG 3 %%%%%%%%%%%%%

% PLEASE SEE BELOW UPDATE July 2020

%\change{In Fig.~\ref{irobonding}, we show the Ir-O hybridization under compressive(a), tensile (c), and no strain (b).  Our DFT %calculations at $5\%$
%compressive strain show that the hybridization of the Ir
%orbitals with oxygen is reduced strongly, suggesting that the relative Ir-O hybridization is indeed
%directly responsible for the resulting suppression of $t_1$ under
%compressive strain.}

%%%%%%%%%%%%%%%% FIG Ap 2%%%%%%%%%%%%%
%\begin{figure}[!h]
%	\centering
%	\includegraphics[width=0.99\linewidth]{pic/irobonding_DFT.png}
%	\caption{
%		\change{The Ir-O hybridization and the change of the bandwidth under (a) compressive, (c) tensile (c), and (b) no strain.}
%		\label{irobonding}
%	}
%\end{figure}
%%%%%%%%%%%%%%%% FIG Ap 2%%%%%%%%%%%%%

%\section*{Supplementary Note 3}
\subsection{}
In the following, we present Slater-Koster analysis of the oxygen-mediated overlap of $xz$ and $yz$ orbitals. The indirect overlap between the two rotated NN $xz$/$yz$ orbitals is mediated via the oxygen $p_z$ orbital and calculated as 
\begin{align}
	\tilde{E}_{\mathrm{xz/yz,z,xz/yz}}&=(\cos(\theta)E_{\mathrm{z,xz/yz}}^{-l,m,n}\mp
    \sin(\theta)E_{\mathrm{z,yz/xz}}^{-l,m,n})\nonumber \\
	&\times(\cos(\theta)E_{\mathrm{z,xz/yz}}^{l,m,n}\pm\sin(\theta)E_{\mathrm{z,yz/xz}}^{l,m,n})/\Delta_{\mathrm{pd}},
	\label{ExzOxz}
\end{align}
where $l=\cos{\theta}$, $m=\sin{\theta}$, $n=0$ are the directional
cosines of the vector connecting the Ir atom to neighboring (along the
$a'$-axis) oxygen O in the units of Ir-O distance $d_{\rm Ir-O}= 0.5
d_0^{\rm Ir-Ir}/\cos{\theta}$.  $E_{\mathrm{z,xz/yz}}$ are the $p-d$ Slater-Koster
integrals~\cite{HarrisonBook2004}, and
    $\Delta_{\mathrm{pd}}=E^{\mathrm{on-site}}_{\mathrm{xz/yz}}-E^\mathrm{{on-site}}_{\mathrm{z}}$ is the difference
between on-site energies of the corresponding $d$ and $p$ orbitals,
which are obtained from the projected density of states and shown in
Supplemental Figure \ref{ExzyzviaO_full}(a).

We obtain that indirect overlap for $yz$ orbitals along $a'$ axis is
zero (not shown), and indirect $xz$-$xz$ overlap decreases slightly
under compressive strain (Supplemental Figure \ref{ExzyzviaO_full}(b)).
The contribution of the indirect $xz$-$xz$ overlap to the overall
$xz$-$xz$ hopping is significantly reduced compared to the case of the
$xy$-$xy$ Ir-Ir overlap by the denominator $\Delta_{\mathrm{pd}}$, the
energy difference between corresponding Ir and O orbitals (Supplemental
Figure \ref{ExzyzviaO_full}(a)). The $xy$ orbital is lower in energy than $xz/yz$ orbitals due to the octahedra elongation, and thus closer to the oxygen orbitals, making denominator $\Delta_{\mathrm{pd}}$ smaller and thus the overall contribution of the indirect hopping bigger for $xy$ orbital. Moreover, upon compressive strain, the octahedra get further elongated~\cite{BhandariNJP2019}, and, therefore, the denominator would
further increase (decrease) for $xz$- ($xy$-) indirect hopping contribution. 
	
%TO CHIEN-CHENG UPDATE July 2020
	
%I don't think we need the following discussion any more as we now have the actual energies by Wei-Chih. I therefor suggest to not include it and also not include bandwidth plot (previously, Fig 10):
	
%Indeed, we have examined the effect of distortions by comparing the relaxed structure (with octahedra elongation) and another unrealistic structure (without elongation) using the same a-,b-, and c-axis. Upon elongation, the contribution of the bands with $xy$ character changes dramatically, with the bandwidth shrinking from roughly from $3.8$ eV to $2.5$ eV, while the bands with $xz$ and $yz$ character only change slightly. We note that overall, the role of oxygens in the strain-induced changes of the electronic structure in iridates and other transitional-metal oxides seems to be underestimated and requires further detailed studies.

%%%%%%%%%%%%%%%% FIG Ap3 %%%%%%%%%%%%%
\begin{figure}[!h]
	\centering
	\includegraphics[width=0.4\linewidth]{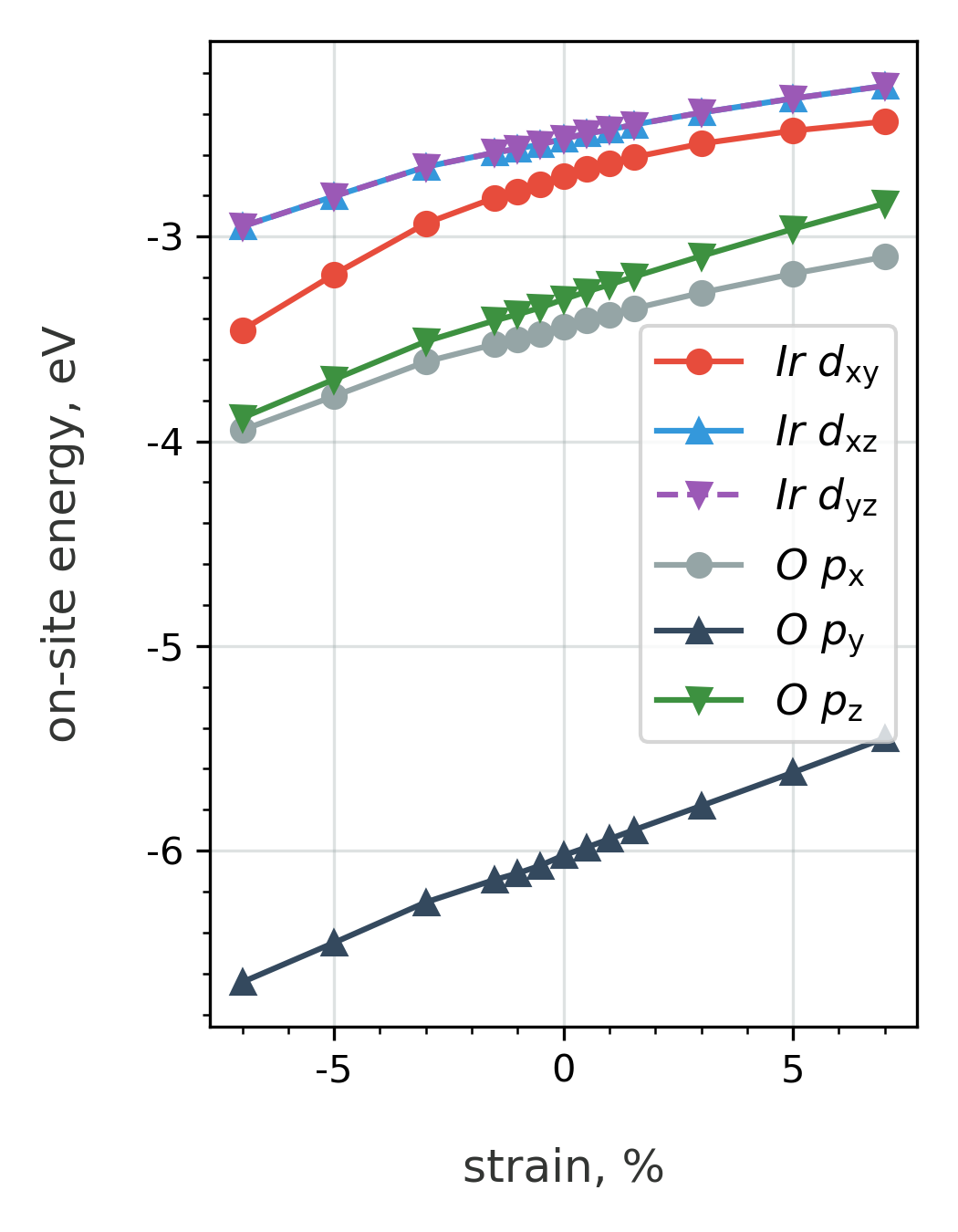}\label{Eonsite}
	\llap{\parbox[b]{2.50in}{\textbf{a}\\\rule{0ex}{3.2in}}}
	\includegraphics[width=0.4\linewidth]{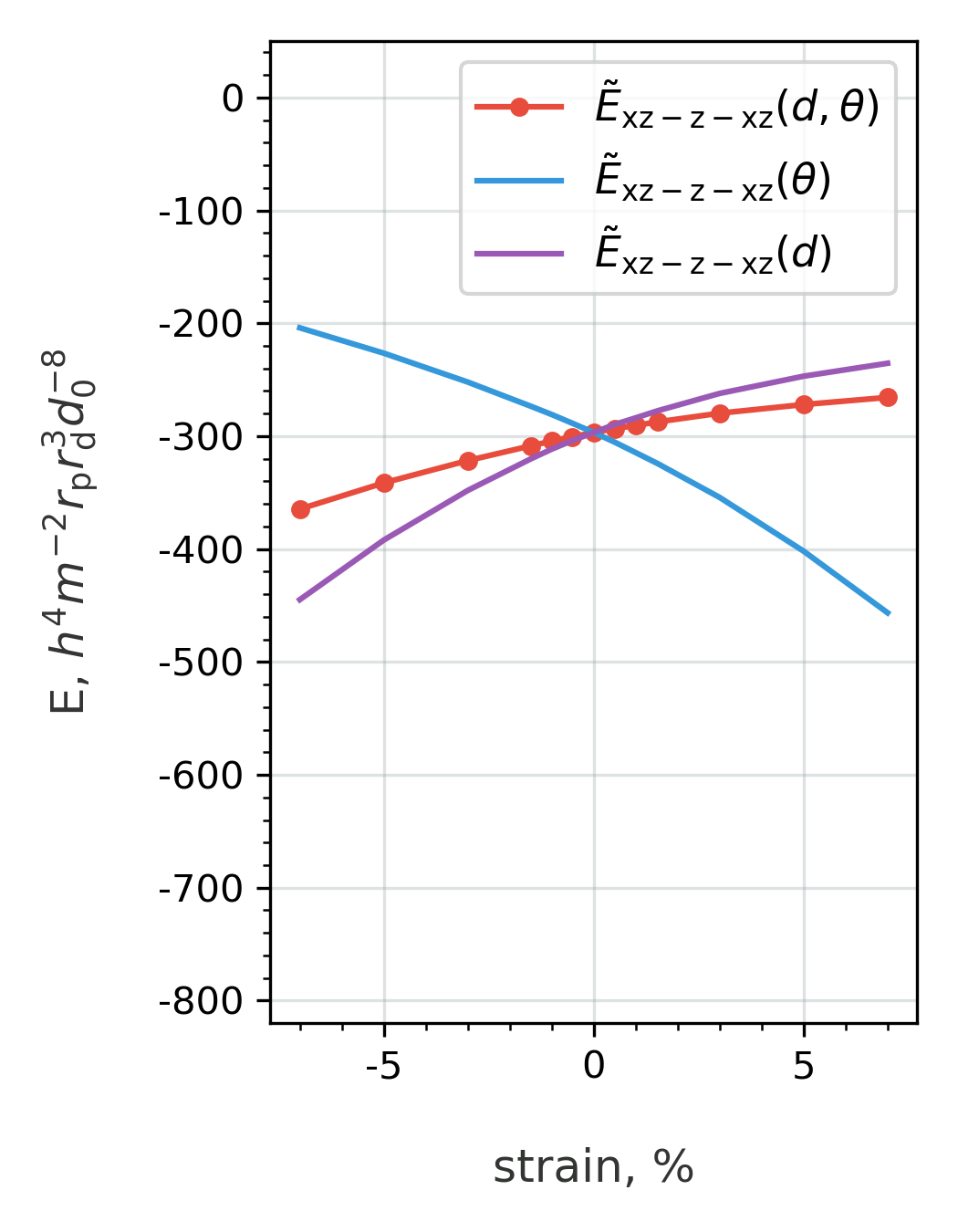}\label{ExzviaOSLaterKoster}\llap{\parbox[b]{2.50in}{\textbf{b}\\\rule{0ex}{3.2in}}}
	\caption{ (a) The strain dependency of the on-site energy of iridium $d$ orbitals and neighboring oxygen $p$ orbitals along $a'$-axis. We note that due to the lattice symmetry, for an oxygen neighboring iridium atom along $b'$-axis, the energies of $p_\mathrm{x}$ and $p_\mathrm{y}$ orbitals are interchanged.  We note that when averaged over the unit cell, the energies of $p_\mathrm{x}$ and $p_\mathrm{y}$ orbitals are the same, as expected from symmetry considerations.(b) Slater-Koster integrals for indirect hopping between rotated Ir NN orbitals $\tilde{E}$ via oxygen $p$ orbitals (along $a'$-axis) as a function of octahedra in-plane rotation angle $\theta$, Ir-Ir distance $d$. 
		%Dashed lines are used throughout the paper to show Wannier function fitting results, while solid lines are used to show results obtained using Slater-Koster and Glazer's analytical approaches.
		\label{ExzyzviaO_full}}
\end{figure}
%%%%%%%%%%%%%%%% FIG Ap3 %%%%%%%%%%%%%

\clearpage
\newpage

%\bibliographystyle{naturemag}
%\bibliography{strained_Ir214_weichih}

\end{document}